\documentclass[12pt,a4paper]{article} 

\usepackage[margin=1in]{geometry} 
\usepackage{amsfonts,amscd,amssymb, mathtools,mathrsfs, dsfont, bbm,bbding} 
\usepackage[amsmath,amsthm,thmmarks]{ntheorem} 
\usepackage{graphicx,xypic,color,float} 
\usepackage{indentfirst}
\usepackage{lmodern}
\usepackage[T1]{fontenc} 
\usepackage{enumerate,listings,verbatim,paralist}
\usepackage{setspace,xspace}
\usepackage[colorlinks=true,citecolor=blue]{hyperref}
\usepackage{extarrows}
\usepackage{pdfpages}
\usepackage{ulem}

\usepackage{setspace}
\AtBeginDocument{%
\addtolength\abovedisplayskip{-0.15\baselineskip}%
\addtolength\belowdisplayskip{-0.15\baselineskip}%
\addtolength\abovedisplayshortskip{-0.15\baselineskip}%
\addtolength\belowdisplayshortskip{-0.15\baselineskip}%
}

\newtheorem{thm}{Theorem} [section]
\newtheorem{prop}[thm]{Proposition}
\newtheorem{coro}[thm]{Corollary}
\newtheorem{lemma}[thm]{Lemma}
\newtheorem{defn}{Definition}[section]
\newtheorem{assmp}{Assumption}[section]
\newtheorem{remark}{Remark}[section]

\numberwithin{equation}{section} 

\renewcommand{\geq}{\geqslant}
\renewcommand{\leq}{\leqslant}

\newcommand{\citethm}[1]{Theorem \ref{#1}}
\newcommand{\citeprop}[1]{Proposition \ref{#1}}
\newcommand{\citecoro}[1]{Corollary \ref{#1}}
\newcommand{\citelem}[1]{Lemma \ref{#1}}

\newcommand{\citeassmp}[1]{Assumption \ref{#1}}
\newcommand{\citeremark}[1]{Remark \ref{#1}}
\newcommand{\citesec}[1]{Section \ref{#1}}
\newcommand{\citefig}[1]{Figure \ref{#1}}

\newcommand{\opfont}{\mathbb}

\newcommand{\BE}[2][]{\ensuremath{\operatorname{\opfont{E}}^{#1}\!\left[#2\right]}}

\newcommand{\R}{\ensuremath{\operatorname{\mathbb{R}}}}

\newcommand{\dd}{\ensuremath{\operatorname{d}\! }}
\newcommand{\dt}{\ensuremath{\operatorname{d}\! t}}
\newcommand{\ds}{\ensuremath{\operatorname{d}\! s}}

\newcommand{\ddp}{\ensuremath{\operatorname{d}\! p}}

\newcommand{\setw}{\mathscr{W}}
\newcommand{\setg}{\mathscr{G}}
\newcommand{\setq}{\mathscr{Q}}
\newcommand{\setr}{\mathscr{R}}
\newcommand{\setc}{\mathscr{C}}

\def\inw{\nu}

\newcommand{\nn}{\nonumber}

\newcommand{\pa}{\mathbf{P}}
\newcommand{\wa}{\beta_{\mathrm{insured}}}

\newcommand{\uinsured}{\mathcal{U}_{\mathrm{insured}}}
\newcommand{\uinsurer}{\mathcal{U}_{\mathrm{insurer}}}

\newcommand{\BV}{\ensuremath{\mathscr{E}}}
\newcommand{\einf}{\ensuremath{\mathrm{ess\:inf\:}}}
\newcommand{\esup}{\ensuremath{\mathrm{ess\:sup\:}}}

\newcommand{\barq}{\overline{Q}}
\newcommand{\ep}{\varepsilon}
\newcommand{\opl}{\Phi}
\newcommand{\inprime}{\varpi}
\newcommand{\inone}{\sigma}
\newcommand{\opp}{\Psi}

\newcommand{\penal}{\mathscr{P}}
\newcommand{\baropl}{\overline{\opl}}
\newcommand{\opll}{\psi}

\newcommand{\ms}{m_0}

\newcommand{\dsp}{\displaystyle}

\begin{document}
\title{Pareto optimal moral-hazard-free insurance contracts \\ in behavioral finance framework }
\author{Zuo Quan Xu\thanks{Department of Applied Mathematics, The Hong Kong Polytechnic University, Kowloon, Hong Kong. Email: \url{maxu@polyu.edu.hk}. The author acknowledges financial support from NSFC (No.11971409), The Hong Kong GRF (No.15204216 and No.15202817), The PolyU-SDU Joint Research Center on Financial Mathematics and the CAS AMSS-POLYU Joint Laboratory of Applied Mathematics, The Hong Kong Polytechnic University.}}
\date{May 2021}
\maketitle

\begin{abstract} 
This paper investigates Pareto optimal (PO, for short) insurance contracts in a behavioral finance framework, in which the insured evaluates contracts by the rank-dependent utility (RDU) theory and the insurer by the expected value premium principle. The incentive compatibility constraint is taken into account, so the contracts are free of moral hazard. The problem is initially formulated as a non-concave maximization problem involving Choquet expectation, then turned into a quantile optimization problem and tackled by calculus of variations method. The optimal contracts are expressed by a double-obstacle ordinary differential equation for a semi-linear second-order elliptic operator with nonlocal boundary conditions. We provide a simple numerical scheme as well as a numerical example to calculate the optimal contracts. Let $\theta$ and $\ms$ denote the relative safety loading and the mass of the potential loss at 0. We find that every moral-hazard-free contract is optimal for infinitely many RDU insureds if $0<\theta<\frac{\ms}{1-\ms}$; by contrast, some contracts such as the full coverage contract are never optimal for any RDU insured if $\theta>\frac{\ms}{1-\ms}$. We also derive all the PO contracts when either the compensations or the retentions loss monotonicity. \bigskip
\end{abstract}

\section{Introduction}
Probability weighting function (also called probability distortion function) (see \cite{TF95,P98}) plays a key role in many behavioral theories of choice under uncertainty, such as Kahneman and Tversky's \cite{KT79,TK92} cumulative prospect theory, Yaari's \cite{Y87} dual model, Lopes' \cite{L87} SP/A model and Quiggin's \cite{Q82} rank-dependent utility (RDU, for short) theory. These behavioral finance theories provide satisfactory explanations of the many paradoxes for which the classical expected utility (EU) theory fails to account (see, e.g. \cite{FS48,A53,E61,MP85}). 
\par
In recent years, much attention has been paid to the theoretical study of behavioral finance models under uncertainty (such as portfolio choice and optimal stopping models) involving probability weighting function; see, e.g., \cite{JZ08,HZ11,JZZ11,XZ13,X14,XZ16,HX16, RV17, W18}. 
A typical approach to such a problem is described as follows. Rather than looking for the optimal strategy directly for the original stochastic control or optimization problem, one first reduces the problem to a corresponding quantile optimization problem, in which the decision variable becomes a quantile function (or simply called a quantile, which is the inverse of a probability distribution function). With this change, solving a stochastic control or optimization problem reduces to solving a static deterministic quantile optimization problem. In the second step, the deduced quantile optimization problem is tackled by deterministic optimization techniques, such as completing the square method (see \cite{HX16}), convex analysis (see \cite{HZ11,XZ13, HJZ15, W18}) and calculus of variations method (see \cite{XZ16}). The last step is to recover the optimal strategy for the original problem by appealing to some proper hedging theories, such as the backward stochastic differential equation theory for portfolio choice problems (see \cite{HZ11}) and the Skorokhod embedding theory for optimal stopping problems (see \cite{XZ13}). As every quantile function is non-decreasing, any quantile optimization problem must take this monotonicity (as a minimum) constraint into consideration. This becomes the main hurdle when tackling many quantile optimization problems. 
\par
Researchers generally tackle quantile optimization problems in isolation and often under fairly strong assumptions (see, e.g., \cite{JZ08, HZ11}) due to lack of a systematic method. Xia and Zhou \cite{XZ16} provided a systematic approach by calculus of variations method. They demonstrated the utility of their approach by solving a portfolio choice problem under the RDU theory. Shortly after, the author \cite{X16} of the present paper introduced an alternative simple method, namely the change of variables and relaxation method, to solve a class of quantile optimization problems including the one considered in \cite{XZ16}. For this class of problems, the constraints on the quantiles are almost minimum: beyond the monotonicity constraint (which is necessary for quantile optimization problems, as previously noted), the only other constraint arising from the models is the so-called budget constraint, which, mathematically speaking, is a one-dimension linear constraint that can be easily dealt by Lagrangian method. 
\par
Probability weighting function is also widely used in the risk-sharing literature. In the context of insurance, the primary risk-sharing problem is to design an insurance contract between insurer and insured that achieves Pareto optimality/efficiency. Although there is plenty of work done on designing optimal compensations under the RDU theory (\cite{CDT00,DS07,CD08,BHYZ15, XZZ19, G19}), most studies assume that the probability weighting function takes a special shape, such as convex, concave or inverse-$S$-shaped. As with the aforementioned investment problems, many insurance contracts problems can also be solved by the same routine, namely translating them into quantile optimization problems, determining the optimal quantiles, and translating back to discern the optimal contracts. 
\par
However, there is a key difference in how quantile optimization problems are formulated for investment models (called ``first-type'') and that for optimal insurance problems (called ``second-type''). When considering insurance problems, one has to take concerns from insurer and insured into account simultaneously; as a consequence, both compensation and retention functions are a prior non-decreasing for the optimal contracts. Both Huberman, Mayers and Smith Jr \cite{HMS83} and Picard \cite{P00} called the non-decreasing condition of compensation and retention the \emph{incentive compatibility constraint} for optimal insurance contracts. Mathematically speaking, this leads to a new class of quantile optimization problems, in which the derivatives of decision quantiles are bounded. This is an infinite-dimension constraint. Bernard et al. \cite{BHYZ15} studied an insurance contract design model under the RDU theory, but ignored the incentive compatibility constraint. As a consequence, their result suffers from a moral hazard issue: providing incentives for the insured to falsely report actual losses. Xu et al. \cite{XZZ19} examined the same model but took the {incentive compatibility constraint} into consideration to avoid this moral hazard issue; but due to technical difficulties, they can only partially solve the problem by imposing restrictive assumptions on the loss and probability weighting function. Ghossoub \cite{G19} revisited the same problem by imposing a state-verification cost that the insurer can incur in order to verify the loss severity, hence automatically ruling out any ex post moral hazard that could otherwise arise from possible misreporting of the loss by the insured. 
\par
To the best of our knowledge, no systematic approach has been developed to solve the second-type quantile optimization problems. Although the calculus of variations method has been used in the insurance literature, most applications did not take the constraint of incentive compatibility into account. For example, Spence and Zeckhauser \cite{SZ71} used this method to solve an optimal insurance problem under the EU theory without considering the constraint. However, because the optimal contract turns out to be the classical deductible, it coincidentally satisfies the compatibility constraint. As mentioned above, when problems are considered in behavioral finance framework (such as the RDU theory), the optimal contracts can lead to moral hazard issue. Xu et al. \cite{XZZ19} is only study we could identify that puts this constraint into the model setup. 
\par
In this paper we investigate the Pareto optimal (PO) insurance contracts in a behavioral finance framework, in which the insured evaluates contracts by the RDU theory and the insurer is risk neutral and using the expected value premium principle. The {incentive compatibility} constraint is taken into account when modeling, so the contracts are free of the aforementioned moral hazard issue. The problem is formulated as a non-concave optimization problem which involves Choquet expectation (see \eqref{opi0}). To solve it, we first translate the problem into a quantile optimization problem of the second-type by change of variables in \citesec{quantileproblem}. Then apply calculus of variations method to get the optimality condition for the latter in \citesec{solution}. Finally we derive all the PO insurance contracts in \citethm{main1} and study their properties in \citesec{property}. In this process, an ordinary integral-differential equation (OIDE) \eqref{vi001} and an ordinary differential equation (ODE) \eqref{vi002} play the key role. We also derive all the PO insurance contracts when either the compensations or the retentions are not required to be non-decreasing in \citesec{nonmonotone}.

\par
The main mathematical contribution of this paper is to express the solution by a double-obstacle ODE for a {semi-linear} second-order elliptic operator with nonlocal boundary conditions (\citethm{main1}). At first sight, the ODE \eqref{vi001} looks like a standard double-obstacle problem in the financial economics literature, however, the obstacles are put on the highest order gradient of the unknown function; by contrast, they are put on the lower order gradients in the literature. To the best of our knowledge, it is the first time that this type of double-obstacle problems appears in the financial economics literature. We prove the existence and uniqueness of the solution to the ODE \eqref{vi002} as well as the OIDE \eqref{vi001} from the pure optimization point view. Furthermore, we also provide a numerical scheme to calculate all the PO contracts in \citesec{numerical}. The problem is thus completely solved. We discover that the optimal solution obtained in \cite{XZ16} and \cite{X16} can be expressed by a {single-obstacle} problem for a {linear} second-order elliptic operator. Last, if the incentive compatibility constraint is ignored when modeling, namely either the compensations or the retentions are not required to be non-decreasing, we also derive all the PO contracts. In the former case, all the PO contracts are given explicitly via a free parameter and the concave envelope of a known function (see \citethm{main3}). Moreover, two equivalent conditions for these contracts to be ex post moral-hazard-free are provided in \citeremark{remarkmain3} and \citeremark{remarkmain4}.
\par
The main economic contributions of this paper are as follows. First, we reveals in \citeprop{EUcase} the classical assertion that all the PO contracts are of deductible type in the EU theory framework (see, e.g. \cite{SZ71}). Second, we give an equivalent condition under which the PO contracts are of deductible type in \citethm{special1}. Next, we find that the relative safety loading $\theta$ of the insurer and the mass $\ms$ of the potential loss at 0 play the key role in determining the type of the PO contracts. In fact, if the relative safety loading is small or the mass $\ms$ is big, namely $0<\theta<\frac{\ms}{1-\ms}$, then every moral-hazard-free contract will be accepted by infinitely many RDU insureds (see \citethm{main2}); by contrast, if the relative safety loading is big or the mass $\ms$ is small, namely $\theta>\frac{\ms}{1-\ms}$, some contracts (such as the full coverage contract) will be rejected by every RDU insured (see \citecoro{coro1}). Last, we find that some PO contracts may not be moral-hazard-free if the incentive compatibility constraint is ignored when modeling. 
\par
The rest of this paper is organized as follows. In \citesec{problem}, we introduce a PO insurance problem. In \citesec{quantileproblem}, the problem is turned into a concave quantile optimization problem via change of variables. \citesec{solution} is devoted to solving the quantile optimization problem by calculus of variations method and providing a numerical scheme with an example to calculate the PO contracts. In \citesec{property}, we discuss properties of the PO contracts. \citesec{nonmonotone} studies two models where the compensations or the retentions are not required to be non-decreasing. \citesec{conclusion} concludes the paper. 

\subsection*{Notation}
Throughout the paper, we fix an atom-less probability space. 
For any random variable $Y\geq 0$, we denote its probability distribution function by $F_{Y}$; and define its quantile function (or the left-continuous inverse function of $F_{Y}$) by
\[F^{-1}_{Y}(p):=\inf\big\{z\geq 0\;\big|\; F_{Y}(z)\geq p\big\},\quad p\in(0,1],\]
with the convention that $\inf\emptyset=+\infty$. We also set $F^{-1}_{Y}(0):=\lim_{p\to 0+}F^{-1}_{Y}(p)$. By this definition, $F^{-1}_{Y}(0)=\einf Y$ and $F^{-1}_{Y}(1)=\esup Y$. So $F^{-1}_{Y}(1)<\infty$ if and only if $Y$ is upper bounded. 
\par
All the quantile functions by definition are nonnegative, non-decreasing and left-continuous.
However, they may not be continuous in general. In this paper, we will mainly deal with absolutely continuous quantiles due to the incentive compatibility constraint involved.
\par
We denote by $C^{2-}([0,1])$ the set of functions $f: [0,1]\to \R$ which are differentiable and their derivatives $f'$ are absolutely continuous functions on $[0,1]$. Clearly $C^{2}([0,1])\subsetneqq C^{2-}([0,1])\subsetneqq C^{1}([0,1])$. 
\par
In our argument, ``almost surely'' and ``almost everywhere'' (a.e.) may be suppressed for notation simplicity in some circumstances when no confusion occurs.

\section{Problem formulation}\label{problem}

In Pareto optimal (PO, for short; also called Pareto efficient) insurance problem, one seeks the best way for the insurer (an insurance company) and the insured (``She'') to share a potential loss to achieve Pareto optimality or efficiency.
\par
We use random variable $X\geq 0$ to denote potential loss that is covered by the insurance contract.
Let $I(x)$ be the loss borne by the insurer when a real loss $x$ occurs. It is called the \emph{compensation} (or \emph{indemnity}) function in the insurance literature. A compensation is called full coverage if $I(x)\equiv x$; called deductible (with deductible $d$) if $I(x)\equiv \max\{x-d,0\}$. Let $\setc$ denote the set of acceptable compensations that will be specified shortly. 
\par
Following the standard insurance literature, an insurance contract is a pair $(\pa, I)$, where $\pa\in\R$ is a premium that the insured pays to the insurer at initial time and $I\in \setc$ is an acceptable compensation. Suppose the insured and the insurer evaluate the contract as $\uinsured(\pa, I)$ and $\uinsurer(\pa, I)$, respectively. We say a contract $(\pa, I)$ is Pareto optimal/efficient if there is no other feasible contract $(\pa', I')$ such that 
\[\uinsured(\pa', I')\geq \uinsured(\pa, I), \quad \uinsurer(\pa', I')\geq \uinsurer(\pa, I)\] 
and 
\[\uinsured(\pa', I')+\uinsurer(\pa', I')> \uinsured(\pa, I)+\uinsurer(\pa, I).\] 
In other words, it is impossible to improve one of the insured's and insurer's valuations without reducing the other one's. All the PO contracts form a set, called Pareto frontier. 
\par
Economically speaking, the insured should be happier if she pays less premium for the same compensation or receives a higher compensation without paying more premium, so we assume $\uinsured(\pa, I)$ is decreasing in $\pa$ and increasing in $I$. Similarly, we assume $\uinsurer(\pa, I)$ is increasing in $\pa$ and decreasing in $I$. Also assume both $\uinsured(\pa, I)$ and $\uinsurer(\pa, I)$ are continuous in $\pa$. It is an easy exercise to show that a contract $({\pa}^*, {I}^*)$ is PO if and only if there exists $\gamma\in\R$ such that $({\pa}^*, {I}^*)$ is an optimal solution to the problem
\begin{align} \label{pe0}
\sup_{\pa\in\R, \; I\in\setc} &\quad \uinsured(\pa, I) \\
\mathrm{s.t.}&\quad \uinsurer(\pa, I)\geq \gamma. \nn
\end{align} 
The last inequality is often called the participation constraint for the insurer. If we runs through all the values of $\gamma$, we will get all the PO contracts.
\par
Variations in risk measures lead to different PO insurance contracts models. We need to specify our model for further analysis. We start with the set of acceptable compensations $\setc$. Let $R(x)$ be the loss borne by the insured when a real loss $x$ occurs. It is called the \emph{retention} function. Let $\setr$ denote the set of the retention functions. 
Mathematically speaking, one always has 
\begin{align} \label{totalloss}
I(0)=R(0)=0,\quad I(x)+R(x)=x,\quad x\geq 0.
\end{align} 
Therefore, $\setr=\{\cdot-I(\cdot): I\in\setc\}$. On the other hand, economically speaking, both the insurer and the insured should bear a greater financial responsibility when a bigger loss occurs. 
For, if one party borne less responsibility, it could result in the moral hazard issue (see more in \cite{XZZ19}). Therefore, mathematically speaking, one also requires that 
\begin{align} \label{monotonloss}
I(x)\geq I(y),\quad R(x)\geq R(y),\quad x\geq y\geq 0.
\end{align} 
This is called the \emph{incentive compatibility constraint} (see, e.g. \cite{HMS83,P00}). In our model, we request all the compensations and retentions should satisfy the above constraints \eqref{totalloss} and \eqref{monotonloss}. We call them \emph{moral-hazard-free}. For instance, all deductible and proportional coverage compensations (which satisfy $I(x)\equiv cx$ for some $0<c<1$) are moral-hazard-free. Cleary we can combine the constraints \eqref{totalloss} and \eqref{monotonloss} into the following one:
\begin{align*} 
I(0)=0,\quad 0\leq I(x)-I(y)\leq x-y, \quad x\geq y\geq 0.
\end{align*} 
In another words, we have the classical result of Denneberg, 
\begin{align*} 
\setc&=\Big\{I:[0,\infty)\to [0,\infty)\; \big|\; \mbox{$I$ is absolutely continuous }\nn\\
&\qquad\quad~\mbox{ with $I(0)=0$ and $0\leq I'\leq 1$ almost everywhere (a.e.).}\Big\}.
\end{align*} 
Recall $\setr=\{\cdot-I(\cdot): I\in\setc\}$, so the above implies  
\begin{align*} 
\setr &=\Big\{R:[0,\infty)\to [0,\infty)\; \big|\; \mbox{$R$ is absolutely }\nn\\
&\qquad\quad~\mbox{ continuous with $R(0)=0$ and $0\leq R'\leq 1$ a.e.}\Big\}.
\end{align*} 
Clearly $\setc=\setr$ in this case. 
Because the derivative of the compensation function $I$ (as well as the retention function $R$) is bounded in $[0, 1]$, on one hand, economically it avoids the potential moral hard issue; on the other hand, mathematically it also makes the related PO moral-hazard-free insurance contracts hard to find since it is an infinity-dimension constraint. 
\par
It is yet to specify $\uinsured(\pa, I)$ and $\uinsurer(\pa, I)$. Following the existing literature (see, e.g, \cite{A63,R79,GS96}), we assume the insurer is risk neutral and evaluates contracts by the expected value premium principle, that is 
\begin{align} 
\uinsurer(\pa, I)=\pa-(1+\theta)\BE{I(X)}, 
\end{align} 
where $\theta>0$ is a constant, called the \emph{relative safety loading}. The value of $\theta$ will have a significant impact on the type of the PO contracts. Meanwhile, we assume the insured evaluates contracts by
\begin{align} 
\uinsured(\pa, I)=\BV\big(\wa-\pa-X+I(X)\big), 
\end{align} 
where $\wa$ is a constant standing for the initial wealth of the insured. Note $\wa-\pa-X+I(X)$ is the insured's net wealth after claim. The risk measure $\BV$ will be specified shortly. 
\par
Under the above specific setting, problem \eqref{pe0} becomes 
\begin{align*} 
\sup_{\pa\in\R, \; I\in\setc}&\quad \uinsured(\pa, I)=\BV\big(\wa-\pa-X+I(X)\big) \\
\mathrm{s.t.}&\quad \pa-(1+\theta)\BE{I(X)}\geq \gamma. 
\end{align*} 
Recall that $\uinsured(\pa, I)$ is assumed to be decreasing in $\pa$, so any PO contract $(\pa^*, I^*)$ shall make the constraint tight, namely
\[\pa^*=\gamma+(1+\theta)\BE{I^*(X)}.\]
Therefore, it suffices to study the problem 
\begin{align} \label{opi0}
\sup_{I\in\setc}&\quad \BV\big(\wa-\gamma-(1+\theta)\BE{I(X)}-X+I(X)\big).
\end{align} 
\par
Let $R^*_{\beta}$ denote an optimal solution to the following problem, 
\begin{align} \label{opi}
\sup_{R\in\setr}&\quad \BV\big(\beta+(1+\theta)\BE{R(X)}-R(X)\big).
\end{align} 
Then $I^*(x):=x-R^*_{\wa-\gamma-(1+\theta)\BE{X}}(x)$ is an optimal solution to problem \eqref{opi0}, and 
\[\big(\gamma+(1+\theta)\BE{X-R^*_{\wa-\gamma-(1+\theta)\BE{X}}(X)}, \cdot-R^*_{\wa-\gamma-(1+\theta)\BE{X}}(\cdot)\big)\] 
is a PO contract. In fact it is easily seen that every PO contract is of the form
\[\big(\wa-\beta-(1+\theta)\BE{R^*_{\beta}(X)}, \cdot-R^*_{\beta}(\cdot)\big)\] 
for some $\beta$. When $\beta$ runs through all admissible values, we will get all the PO contracts. We notice that even for the same optimal retention function $R^*_{\beta}$, different insurers with different initial values of $\wa$ would prefer to pay different premiums. 
The optimal premium linearly grows with respect to the insurer's initial value. 
\par
Now our problem reduces to solving \eqref{opi}. Without confusion, we also call its solution (which is indeed an optimal retention) a PO moral-hazard-free contract. A PO contract is called deductible if the compensation in the contract is a deductible one. 
\par
In insurance practice, insurer usually buys a reinsurance contract to cover losses above certain threshold, so the insurer cares only about losses up to this threshold. Because of this, similar to \cite{XZZ19}, we put the following technical assumptions on $X$. 
\begin{assmp}\label{ass1}
The quantile function $F_{X}^{-1}$ of the potential loss $X$ satisfies $F_{X}^{-1}(0)=\einf X=0$ and $F_{X}^{-1}(1)=\esup X<\infty$. Furthermore, it is absolutely continuous on $[0,1]$ and $\big(F_{X}^{-1}\big)'(p)> 0$ for a.e. $p\in(\ms,1)$, where $\ms:=F_X(0)<1$. 
\end{assmp} 
\citeassmp{ass1} allows $X$ to have a positive mass $\ms$ at 0, which is the most common and important case in insurance practice. We will show that the value of $\ms$ has a significant impact on the type of the PO contracts. By this assumption $X$ is essentially bounded, so our subsequent arguments only deal with bounded random variables. Clearly the probability distribution function $F_X$ of $X$ is continuous on $[0,1]$ and strictly increasing on $[\ms,1]$. Moreover, $F_{X}^{-1}(F_X(x))=x$ for all $\einf X\leq x\leq \esup X$, $F_{X}^{-1}(p)=0$ for $p\leq \ms$ and $F_{X}^{-1}(p)>0$ for $p>\ms$. These facts may be used in the subsequent analysis without claim. 
\par
To avoid bankruptcy, we only consider $I\in\setc$ that satisfies the following assumption. 
\begin{assmp}\label{ass2}
We have $\wa-\gamma-(1+\theta)\BE{I(X)}-X+I(X)\geq 0$ in problem \eqref{opi0}.
\end{assmp}
This assumption is equivalent to 
\begin{align} \label{bigbeta}
\beta+(1+\theta)\BE{R(X)}-R(X)\geq 0
\end{align} 
in problem \eqref{opi}. This condition is also applied to subsequent analysis where $\beta$ and $R\in\setr$ are involved. We will not emphasis this again. Now let us introduce the risk measure $\BV$ for the insured in details. In this paper, we consider a behavioral insured, who uses a RDU risk measure $\BV$. The risk measure $\BV$ for a nonnegative random variable $Y$ is defined by 
\begin{align} \label{vdef}
\BV(Y):=\int_{[0,\infty)}u(z)w'(1-F_{Y}(z))\dd F_{Y}(z),
\end{align} 
where $u$ is a utility function which is in $C^2([0,\infty))$ with $u'>0$ and $u''<0$, and $w$ is a probability weighting function in the set 
\begin{align*} 
\setw&:=\Big\{w:[0,1]\to [0,1]\; \big|\; \mbox{$w(p)=\int_0^p f(t)\dt$ for some measurable }\\
&\qquad\quad~\mbox{ function $f: [0,1]\to (0,\infty)$ such that $\int_0^1 f(t)\dt=1$ }\Big\}.
\end{align*} 
Clearly $w$ is continuous and strictly increasing, so it has an inverse function, denoted by $\inw$. Note that $\inw\in\setw$ too.

\begin{remark}
If \citeassmp{ass2} failed, then the net wealth of the insured after claim, $\wa-\gamma-(1+\theta)\BE{I(X)}-X+I(X)$, may be negative. Consequently, the risk preference \eqref{vdef} would be no more suitable to measure it, so a general RDU risk preference, which covers negative positions, should be used. The related problem becomes complicated because the insured may use different probability weighting functions for positive and negative wealth positions. Similarly, one may also consider random wealth $\wa$, in which case the joint distribution of $(\wa, X)$ makes the related contract design problem challenging. These problems are certainly out of the scope of this paper. 
\end{remark}

\begin{remark} \label{remarkxzz}
Xu, Zhou and Zhuang \cite{XZZ19} solved an optimal compensation design problem under the following additional assumptions. First, the probability weighting function $w$ is inverse-$S$-shaped, that is, it is strictly concave on $[0,a]$ and strictly convex on $[a,1]$ for some $0<a<1$. Second, the function $\frac{u''}{u'}$ is non-decreasing. Third, $\frac{w''}{w'}(p)<\frac{u''}{u'}\big(\wa-\gamma-(1+\theta)\BE{X}-F_{X}^{-1}(p)\big)\big(F_{X}^{-1}\big)'(p)$ for $p\in[0,a]$. Clearly these assumptions restrict the applications of their results. One can solve their problem by our method for general cases. We encourage the interested readers to do it. 
\end{remark}

\section{Quantile optimization problem}\label{quantileproblem}
The probability weighting function $w$ in the definition \eqref{vdef} of the risk preference $\BV$ makes the preference a \emph{nonlinear} expectation (it is indeed a Choquet expectation), so \eqref{opi} is a challenging non-concave optimization problem. Since the objective function of problem \eqref{opi} is law-invariant, we can adopt the so-called quantile optimization method to tackle it; see \cite{S04, CD06, CD08, CD11, HZ11, HJZ15, HX16, XZ13, XZ16, X16, W18} for the applications of this method. \par
Our first step to tackle problem \eqref{opi} is to make a change of variable to find an equivalent \emph{concave} quantile optimization problem to it. This clearly will reduce the difficulty of solving it. 
\par 
As is well-known, by change of variable, we can rewrite the risk measure $\BV$ defined by \eqref{vdef} as 
\begin{align}\label{v(y)}
\BV(Y)=\int_{0}^{1}u
\left(F_{Y}^{-1}(p)\right)w'(1-p)\ddp, 
\end{align} 
where $F_{Y}^{-1}$ denotes the quantile function of $Y$. 
\begin{remark}
The risk measure $\BV$ defined by \eqref{vdef} is applicable to nonnegative random variables only. If we use \eqref{v(y)} instead as the definition of the risk measure $\BV$ in our model, then it is applicable to any random variable such that the integral is well defined. Our subsequent analysis still works after necessary adjustment to take care of issues such as integrability when the random variable is unbounded. 
\end{remark}

Next, because our probability space is atom-less, there exists a random variable $\xi$, which is uniformly distributed on $(0,1)$, such that $X=F_{X}^{-1}(\xi)$ almost surely. 
Let 
\begin{align} \label{GR}
G(p):=R\big(F_{X}^{-1}(p)\big),\quad p\in[0,1].
\end{align}
Then it is a non-decreasing function and satisfies \[G(\xi)=R\big(F_{X}^{-1}(\xi)\big)=R(X),\] 
so
\[\BE{R(X)}=\BE{G(\xi)}=\int_{0}^{1}G(t)\dt.\]
Write 
\begin{align*} 
Y&=\beta+(1+\theta)\BE{R(X)}-R(X)\\
&=\beta+(1+\theta)\int_{0}^{1}G(t)\dt-G(\xi).
\end{align*} 
By virtue of the last expression, it is easy to verify that the quantile function of $Y$ is given by 
\[F_{Y}^{-1}(p)=\beta+(1+\theta)\int_{0}^{1}G(t)\dt-G(1-p),\quad \mbox{a.e.}\; p\in[0,1].\]
Taking it into \eqref{v(y)}, we get 
\begin{align*} 
&\quad\;\:\BV\big(\beta+(1+\theta)\BE{R(X)}-R(X)\big)\\
&=\int_{0}^{1}u\Big(\beta+(1+\theta)\int_{0}^{1}G(t)\dt-G(1-p)\Big)w'(1-p)\ddp\\
&=\int_{0}^{1}u\Big(\beta+(1+\theta)\int_{0}^{1}G(t)\dt-G(p)\Big)w'(p)\ddp.
\end{align*} 
\par
We now rewrite the compatibility constraint on $R\in\setr$ in terms of the new decision variable $G$. It is not hard to show $R\in\setr$ if and only if $G\in\setg$\footnote{For more details we refer to \cite{XZZ19}.}, where 
\begin{align*}
\setg&=\Big\{G:[0,1]\to [0,\infty)\;\big|\; \mbox{$G$ is absolutely} \nn\\
&\qquad\quad~\mbox{ continuous with $G(0)=0$ and $0\leq G'\leq h$ a.e.}\Big\}, \label{setg}
\end{align*} 
and 
\begin{align} \label{def:h}
h(p):=\left(F_{X}^{-1}\right)'(p)\geq 0,\quad \mbox{a.e.}\;p\in[0,1].
\end{align} 
Thanks to \citeassmp{ass1}, we have 
\begin{align} \label{h1}
0\leq \int_0^p h(t)\dt=F_{X}^{-1}(p)\leq F_{X}^{-1}(1)=\esup X<\infty.
\end{align} 
After the above change of variables, the study of the optimization problem \eqref{opi} under compatibility constraint reduces to that of the following second-type quantile optimization problem 
\begin{align} \label{opi2}
\sup_{G\in\setg}&\;	\int_{0}^{1}u\Big(\beta+(1+\theta)\int_{0}^{1}G(t)\dt-G(p)\Big)w'(p)\ddp.
\end{align}
In this problem, the objective functional is concave with respect to the decision variable $G$ and the constraint set $\setg$ is convex, so it is a concave optimization problem, which is expected easier to study than the non-concave optimization problem \eqref{opi}. 
\par
Our first result is about the existence and uniqueness of the solution to problem \eqref{opi2}. 
\begin{lemma}\label{lemexist1}
Problem \eqref{opi2} admits a unique optimal solution. 
\end{lemma}

\begin{proof}
By virtue of the Arzel\`a–Ascoli theorem, the existence can be proved by a standard compact argument. The uniqueness is due to the strictly concavity of $u$ and the positivity of $w'$. We leave the details to the interested readers. 
\end{proof}

\begin{remark}\label{G=0}
By the definition \eqref{GR} and \citeassmp{ass1}, we have $G(p)=R\big(F_{X}^{-1}(p)\big)=R(0)=0$ for $p\leq \ms$. {This can also be seen from the constraint $\setg$. In fact, $h(p)=\left(F_{X}^{-1}\right)'(p)=0$ for $p\leq \ms$, so $G'(p)=0$ for a.e. $p\leq \ms$. Consequently $G$ is a constant on $[0,\ms]$, which must be zero as $G(0)=0$.} By virtue of this fact, we can see the objective in \eqref{opi2} can be written as 
\begin{align*} 
u\Big(\beta+(1+\theta)\int_{\ms}^{1}G(t)\dt\Big)w(\ms)
+\int_{\ms}^{1}u\Big(\beta+(1+\theta)\int_{\ms}^{1}G(t)\dt-G(p)\Big)w'(p)\ddp.
\end{align*}
Therefore, the optimal solution to \eqref{opi2} only depends the shape of $w$ on $[\ms,1]$. 
\end{remark}

Following the change of variable argument in \cite{X16}, we further simplify problem \eqref{opi2} to remove $w$ from the objective functional. 
\par 
Recall that $\inw$ is the inverse of $w$ so that $w(\inw(p))\equiv p$. It thus follows
\begin{align} \label{winversew}
w'(\inw(p))\inw'(p)=1, \quad \mbox{a.e.}\; p\in[0,1].
\end{align}
Let
\begin{align} \label{QG}
Q(p) &:=G(\inw(p)),\quad p\in[0,1].
\end{align}
Then by virtue of \eqref{winversew}, 
\begin{align*}
&\quad\;\int_{0}^{1}u\Big(\beta+(1+\theta)\int_{0}^{1}G(t)\dt-G(p)\Big)w'(p)\ddp\\
&=\int_{0}^{1}u\Big(\beta+(1+\theta)\int_{0}^{1}G(\inw(t))\dd \inw(t)-G(\inw(p))\Big)w'(\inw(p))\dd \inw(p)\\
&=\int_{0}^{1}u\Big(\beta+(1+\theta)\int_{0}^{1}Q(t) \inw'(t)\dt-Q(p)\Big)w'(\inw(p)) \inw'(p)\ddp\\
&=\int_{0}^{1}u\Big(\beta+(1+\theta)\int_{0}^{1}Q(t) \inw'(t)\dt-Q(p)\Big)\ddp.
\end{align*} 
It is clearly that $G$ is non-decreasing and left-continuous if and only if so is $Q$. Moreover, since $\inw'>0$, we see that $G'(p) \leq h(p)$ for a.e. $p\in[0,1]$ if and only if $Q'(p)=G'(\inw(p))\inw'(p)\leq h(\inw(p))\inw'(p)$ for a.e. $p\in[0,1]$. Therefore, $G\in\setg$ if and only if $Q\in\setq$, where
\begin{align*}
\setq &:=\Big\{Q:[0,1]\to [0, \infty)\;\big|\; \textrm{ $Q$ is absolutely}\\
&\qquad\quad~\mbox{ continuous with $Q(0)=0$ and $0\leq Q'\leq \hbar$ a.e.}\Big\}, 
\end{align*}
and
\begin{align} \label{hbar1}
\hbar(p):=h(\inw(p)) \inw'(p)\geq 0,\quad \mbox{a.e.}\; p\in[0,1].
\end{align} 
Thanks to \eqref{h1}, 
\begin{align} \label{h2}
\int_{0}^{p}\hbar(t)\dt=\int_{0}^{p}h(\inw(t)) \inw'(t)\dt=F_{X}^{-1}(\inw(p))\leq F_{X}^{-1}(1)=\esup X,\quad p\in[0,1].
\end{align} 
\par
By the above change of variables, solving problem \eqref{opi2} has now reduced to solving the following convex optimization problem 
\begin{align}\label{opi3}
\sup_{Q\in\setq} \int_{0}^{1}u\Big(\beta+(1+\theta)\int_{0}^{1}Q(t) \inw'(t)\dt-Q(p)\Big)\ddp.
\end{align} 
The above change of variables is invertible, so by \citelem{lemexist1} problem \eqref{opi3} also admits a unique solution. 

\begin{lemma}\label{optimals1}
A quantile $\barq$ is the optimal solution to problem \eqref{opi3} if and only if 
\begin{align}
\overline{R}(x)\equiv \barq\big(w(F_X(x))\big)
\end{align}
is the optimal solution to problem \eqref{opi}. 
\end{lemma}

\begin{proof}
By virtue of the change of variables \eqref{GR} and \eqref{QG}, we have the relation 
\begin{align} \label{optimals2}
R(x)\equiv R(F^{-1}_X(F_X(x)))\equiv G(F_X(x))\equiv Q(\inw^{-1}(F_X(x)))\equiv Q(w(F_X(x))).
\end{align}
This completes the proof. 
\end{proof}

\begin{remark}\label{G=01}
Thanks to \citeremark{G=0}, for any $Q\in\setq$, we have $Q(p)=0$ for $p\in[0,w(\ms)]$.
\end{remark}
\par
It follows from \eqref{h2} that, for any $Q\in\setq$, 
\begin{align*} 
0=Q(0)\leq Q(1)=\int_{0}^{1}Q'(p)\ddp\leq \int_{0}^{1}\hbar(t)\dt=F_{X}^{-1}(1)=\esup X<\infty. 
\end{align*} 
Hence the constraint set $\setq$ is a compact set under supreme normal. So we will not emphasis issues such as integrability and boundedness in the subsequent argument.

\section{Optimal solution}\label{solution}
Problem \eqref{opi3} is a second-type quantile optimization problem, in which the derivatives of decision quantiles are both lower and upper bounded. The existing change of variables and relaxation method introduced in \cite{X16} can only deal with the first-type problems, in which the derivatives of decision quantiles are merely lower bounded. It cannot be applied to solve the second-type problems. In this section, we apply calculus variations method to solve problem \eqref{opi3}. 
\par
Because $u$ is concave, problem \eqref{opi3} is a concave optimization problem. So the calculus variations method is expected to provide a not only necessary but also sufficient optimality condition. 
One would need to do extra analysis if a non-concave utility function (such as $S$-shaped utility in Kahneman and Tversky's \cite{KT79,TK92} cumulative prospect theory) would be considered in the model. 

\begin{lemma}[Optimality condition I]\label{op1}
Suppose $\barq\in\setq$. Then $\barq$ is the optimal solution to problem \eqref{opi3} if and only if it satisfies 
\begin{multline} \label{optimalcondition}
\int_{0}^{1}u'\Big(\beta+(1+\theta)\int_{0}^{1}\barq(t) \inw'(t)\dt-\barq(p)\Big) \\
\qquad\times\bigg[\Big((1+\theta)\int_{0}^{1}Q(t) \inw'(t)\dt-Q(p)\Big)-\Big((1+\theta)\int_{0}^{1}\barq(t) \inw'(t)\dt-\barq(p)\Big)\bigg]\ddp\leq 0\qquad \quad\\
\mbox{for any $Q\in\setq$.} 
\end{multline} 
\end{lemma}

\begin{proof}
Suppose $\barq$ is the optimal solution to problem \eqref{opi3}. For any $Q\in\setq$, $\ep\in(0,1)$, define 
\[Q_{\ep}(p)=\barq(p)+\ep (Q(p)-\barq(p)),\quad p\in[0,1].\]
Then it is easy to see $Q_{\ep}\in\setq$. 
By virtue of that $\barq$ is the optimal solution to problem \eqref{opi3} and applying Fatou's lemma, we get 
\begin{align*} 
0 &\geq \liminf_{\ep\to 0+}\frac{1}{\ep}\bigg[\int_{0}^{1}u\Big(\beta+(1+\theta)\int_{0}^{1}Q_{\ep}(t) \inw'(t)\dt-Q_{\ep}(p)\Big)\ddp\\
&\qquad\qquad\qquad-\int_{0}^{1}u\Big(\beta+(1+\theta)\int_{0}^{1}\barq(t) \inw'(t)\dt-\barq(p)\Big)\ddp\bigg]\\
&\geq \int_{0}^{1}u'\Big(\beta+(1+\theta)\int_{0}^{1}\barq(t) \inw'(t)\dt-\barq(p)\Big)\\
&\qquad\times\bigg[\Big((1+\theta)\int_{0}^{1}Q(t) \inw'(t)\dt-Q(p)\Big)-\Big((1+\theta)\int_{0}^{1}\barq(t) \inw'(t)\dt-\barq(p)\Big)\bigg]\ddp.
\end{align*} 
so \eqref{optimalcondition} holds. 
\par
On the other hand side, suppose $\barq\in\setq$ satisfies \eqref{optimalcondition}. 
Because $u$ is concave, we have the elementary inequality $u(y)-u(x)\leq u'(x)(y-x)$ for any $x$, $y\in\R$. So 
\begin{align*} 
&\quad\; u\Big(\beta+(1+\theta)\int_{0}^{1}Q(t) \inw'(t)\dt-Q(p)\Big) -u\Big(\beta+(1+\theta)\int_{0}^{1}\barq(t) \inw'(t)\dt-\barq(p)\Big) \\
&\leq u'\Big(\beta+(1+\theta)\int_{0}^{1}\barq(t) \inw'(t)\dt-\barq(p)\Big)\\
&\qquad\times\bigg[\Big((1+\theta)\int_{0}^{1}Q(t) \inw'(t)\dt-Q(p)\Big)-\Big((1+\theta)\int_{0}^{1}\barq(t) \inw'(t)\dt-\barq(p)\Big)\bigg]
\end{align*} 
for any $Q\in\setq$. 
Integrating both sides, it follows
\begin{align*} 
&\quad\; \int_{0}^{1}\bigg[u\Big(\beta+(1+\theta)\int_{0}^{1}Q(t) \inw'(t)\dt-Q(p)\Big)\\
&\quad\quad\quad\;-u\Big(\beta+(1+\theta)\int_{0}^{1}\barq(t) \inw'(t)\dt-\barq(p)\Big)\bigg]\ddp \\
&\leq \int_{0}^{1}u'\Big(\beta+(1+\theta)\int_{0}^{1}\barq(t) \inw'(t)\dt-\barq(p)\Big)\\
&\qquad\times\bigg[\Big((1+\theta)\int_{0}^{1}Q(t) \inw'(t)\dt-Q(p)\Big)-\Big((1+\theta)\int_{0}^{1}\barq(t) \inw'(t)\dt-\barq(p)\Big)\bigg]\ddp.
\end{align*} 
The right hand side is nonpositive by \eqref{optimalcondition}, so is the left hand side. Therefore, $\barq$ is an optimal solution to problem \eqref{opi3}. 
\end{proof}
\par
By this result it suffices to find a $\barq\in\setq$ to satisfy the condition \eqref{optimalcondition}. It is, however, very hard, if not impossible, to verify \eqref{optimalcondition} since one has to compare $\barq$ with all the other quantiles in $\setq$, which is, intuitively speaking, of the same level of difficulty as problem \eqref{opi3}. 
\par
Our next step is to find an equivalent condition to \eqref{optimalcondition} that can be easily verified. To this end, write 
\begin{align*}
\opl(p)=-\int_{p}^{1}u'\Big(\beta+(1+\theta)\int_{0}^{1}\barq(t) \inw'(t)\dt-\barq(s)\Big)\ds. 
\end{align*}
Then the inequality in \eqref{optimalcondition} is equivalent to 
\begin{align*} 
\int_{0}^{1} \opl'(p)\bigg[\Big((1+\theta)\int_{0}^{1}Q(t) \inw'(t)\dt-Q(p)\Big)-\Big((1+\theta)\int_{0}^{1}\barq(t) \inw'(t)\dt-\barq(p)\Big)\bigg]\ddp\leq 0.
\end{align*} 
Applying integration by parts to the integral and thanks to $\opl(1)=0$ and $Q(0)=\barq(0)=0$, the left hand side in above becomes 
\begin{align*} 
-\opl(0) (1+\theta)\int_{0}^{1}\big(Q(t)-\barq(t)\big) \inw'(t)\dt
+\int_{0}^{1} \opl(p) \big(Q'(p)-\barq'(p)\big)\ddp. 
\end{align*} 
Applying integration by parts to the first integral in above and by virtue of $Q(0)=\barq(0)=0$, the above is equal to 
\begin{align*} 
-\opl(0) (1+\theta)\int_{0}^{1}(1-\inw(t))\big(Q'(t)-\barq'(t)\big)\dt
+\int_{0}^{1} \opl(p)\big(Q'(p)-\barq'(p)\big)\ddp. 
\end{align*} 
Therefore, \eqref{optimalcondition} is equivalent to the following condition
\begin{align} \label{optimalcondition1-1}
\int_{0}^{1} \big(\opl(p)-\opl(0) (1+\theta)(1-\inw(p))\big)\big(Q'(p)-\barq'(p)\big)\ddp \leq 0\; \mbox{ for any $Q\in\setq$.} 
\end{align} 
This condition is equivalent to saying that the linear functional 
\begin{align} \label{optimalcondition1-2}
Q\mapsto \int_{0}^{1} \big(\opl(p)-\opl(0) (1+\theta)(1-\inw(p))\big)Q'(p)\ddp,\quad Q\in\setq,
\end{align} 
is maximized at $\barq$. Because $0\leq Q'\leq \hbar$ for $Q\in\setq$, we conclude that 
\begin{align} \label{optimalcondition2}
\barq\in\setq \mbox{ and }
\begin{cases}
\barq'(p)=\hbar(p), &\;\text{if } \opl(p)>\opl(0) (1+\theta)(1-\inw(p));\\
\barq'(p)\in[0,\hbar(p)], &\;\text{if } \opl(p)=\opl(0) (1+\theta)(1-\inw(p));\\
\barq'(p)=0, &\;\text{if } \opl(p)<\opl(0) (1+\theta)(1-\inw(p)),
\end{cases}\; \mbox{ for a.e. $p\in[0,1]$.}
\end{align} 
It is easy to check that \eqref{optimalcondition2} also implies \eqref{optimalcondition1-1}, so the condition \eqref{optimalcondition2} is equivalent to \eqref{optimalcondition}. We remark that, in contrast to \eqref{optimalcondition}, the condition \eqref{optimalcondition2} is easy to verify since it only depends on $\barq$ itself. 
\par
Although \eqref{optimalcondition2} is easier to verify, it is still uneasy to find or compute $\barq$ from it. We now express the condition \eqref{optimalcondition2} through an ordinary integral-differential equation (OIDE). 
Later, this OIDE will be further reduced to an ordinary differential equation (ODE) so that we can compute the optimal solution $\barq$. To this end, we first introduce a technical result. 
\begin{lemma}\label{threecases}
Suppose $a,b,c,d$ are real numbers with $b\leq c$. Then 
\[\min\{\max\{a-c, \; d\}, \;a-b\}=0\] if and only if 
\begin{align*}
\begin{cases}
a=c, &\text{if } d<0;\\
a\in[b,c], &\text{if } d=0;\\
a=b, &\text{if } d>0.
\end{cases}
\end{align*}
\end{lemma}
\begin{proof} 
The ``if'' part follows from the following facts. 
\begin{itemize}
\item If $d<0$, then $a=c$, so 
\[\min\{\max\{a-c, \; d\}, \;a-b\}=\min\{\max\{0, \; d\}, \;c-b\}=\min\{0, \;c-b\}=0.\]
\item If $d=0$, then $a\in[b,c]$, so 
\[\min\{\max\{a-c, \; d\}, \;a-b\}=\min\{\max\{a-c, \; 0\}, \;a-b\}=\min\{0, \;a-b\}=0.\]
\item If $d>0$, then $a=b$, so 
\[\min\{\max\{a-c, \; d\}, \;a-b\}=\min\{\max\{b-c, \; d\}, \;0\}=\min\{d, \; 0\}=0.\]
\end{itemize}
We next show the ``only if'' part. 
Notice $c\geq b$, so
\[0=\min\{\max\{a-c, \; d\}, \;a-b\}\geq \min\{\max\{a-c, \; d\}, \;a-c\}=a-c,\]
and
\[0=\min\{\max\{a-c, \; d\}, \;a-b\}\leq \min\{\max\{a-b, \; d\}, \;a-b\}=a-b.\] 
Therefore, we always have $a\in[b,c]$. 
Suppose $d<0$ and $a\neq c$, then $a<c$ and 
\[0=\min\{\max\{a-c, \; d\}, \;a-b\}\leq \max\{a-c, \; d\}<0,\]
a contradiction. Hence $a=c$ if $d<0$. Similarly we can prove $a=b$ if $d>0$. 
\end{proof}

\begin{lemma}[Optimality condition II] \label{op2}
Suppose $\barq: [0,1]\to\R$ is an absolutely continuous function. Then $\barq$ is the optimal solution to problem \eqref{opi3} if and only if it satisfies the following OIDE:
\begin{align} \label{vi001}
\begin{cases}
\min\Big\{\max\big\{\barq'(p)-\hbar(p), \;\opl(0) (1+\theta)(1-\inw(p))-\opl(p)\big\},\;\barq'(p)\Big\}=0, \quad\mbox{a.e.}\;p\in[0,1],\\
\barq(0)=0,
\end{cases}
\end{align} 
where
\begin{align} \label{opldef}
\opl(p)=-\int_{p}^{1}u'\Big(\beta+(1+\theta)\int_{0}^{1}\barq(t) \inw'(t)\dt-\barq(s)\Big)\ds. 
\end{align}
\end{lemma}

\begin{proof} 
This is an immediate consequence of the optimality condition \eqref{optimalcondition2} and \citelem{threecases}. 
\end{proof}
\par
As OIDEs are not easy to solve in general, our next goal is to reduce \eqref{vi001} to an easily solved ODE problem. The following technical result will be critical and used frequently in this process. 
\begin{lemma}\label{tech1}
If $\min\{\max\{a,b\},\;c\}=0$, then $\min\{\max\{ak,b\ell\},cm\}=0$ for any $k,\ell,m>0$, vise versa. 
\end{lemma}

\begin{proof}
Suppose $\min\{\max\{a,b\},\;c\}=0$ and $k,\ell,m>0$. Then clearly $c\geq 0$. 
\begin{itemize}
\item If $c>0$, then $\max\{a,b\}=0$, so $a\leq 0$, $b\leq 0$ and $ab=0$. Hence $ak\leq 0$, $b\ell\leq 0$ and $akb\ell=0$, it follows that $\max\{ak,b\ell\}=0$. Thus, $\min\{\max\{ak,b\ell\},cm\}=\min\{0,cm\}=0$ as $cm>0$.
\item If $c=0$, then $\max\{a,b\}\geq 0$, so $a\geq 0$ or $b\geq 0$. Hence $ak\geq 0$ or $b\ell\geq 0$, it follows that $\max\{ak,b\ell\}\geq 0$. Thus, $\min\{\max\{ak,b\ell\},cm\}=\min\{\max\{ak,b\ell\},0\}=0.$
\end{itemize}
The reverse assertion follows trivially by setting $k=\ell=m=1$. 
\end{proof}
\par
We write $f\in C^{2-}([0,1])$ if $f$ is differentiable and its derivative function $f'$ is absolutely continuous on $[0,1]$. Clearly $C^{2}([0,1])\subsetneqq C^{2-}([0,1])\subsetneqq C^{1}([0,1])$. 
\begin{defn}
A function $f\in C^{2-}([0,1])$ is called \emph{good}, if it satisfies all the following conditions: 
\begin{enumerate}
\item $f$ is convex; 
\item $f'>0$;
\item $f$ is linear on the set $\{p\in[0,1]:f(0) (1+\theta)(1-\ms)-f(p)\geq 0\}$. 
\end{enumerate}
\end{defn}
\par
Now we are ready to present an important ODE with boundary conditions. It will play a key role in solving problem \eqref{opi3}.
\begin{equation} \label{vi002}
\begin{cases}
\min\Big\{\max\big\{ \opl''(p)+\hbar(p)u''\big((u')^{-1}\big(\opl'(p)\big)\big), \;\opl(0) (1+\theta)(1-\inw(p))-\opl(p)\big\},\quad\\
\hfill \opl''(p)\Big\}=0, \quad\mbox{a.e.}\;p\in[0,1], \\
\opl(1)=0, \quad
\beta=(1+\theta)\int_{0}^{1} (u')^{-1}\big(\opl'(t)\big)\inw'(t)\dt-(u')^{-1}\big(\opl'(0)\big) \theta.
\end{cases}
\end{equation} 
Our main result, which states the connection between the above ODE and the optimal solution to problem \eqref{opi3}, is given as follows. 
\begin{thm}[Optimal solution]\label{main1}
We have the following assertions. 
\begin{enumerate}[(1).]
\item 
If $\barq$ is the optimal solution to problem \eqref{opi3}. Then 
\begin{align}\label{opldef3}
\opl(p):=-\int_{p}^{1}u'\Big(\beta+(1+\theta)\int_{0}^{1}\barq(t) \inw'(t)\dt-\barq(s)\Big)\ds
\end{align} 
is a solution to \eqref{vi002} in $C^{2-}([0,1])$. Moreover, $\opl$ is a good function. 

\item 
If $\opl$ is a solution to \eqref{vi002} in $C^{2-}([0,1])$. Then 
\begin{align*}
\barq(p) &:=(u')^{-1}\big(\opl'(0)\big)-(u')^{-1}\big(\opl'(p)\big)
\end{align*}
and
\begin{align}
\overline{R}(x)&:=(u')^{-1}\big(\opl'(0)\big)-(u')^{-1}\big(\opl'\big(w(F_X(x))\big)\big)
\end{align}
are the optimal solutions to problems \eqref{opi3} and \eqref{opi}, respectively. 
\end{enumerate}
As a consequence, \eqref{vi002} admits a unique solution in $C^{2-}([0,1])$ and the solution is a good function. 
\end{thm}

\begin{proof}
\begin{enumerate}[(1).]
\item 
Since $\barq$ is absolutely continuous, we have $\opl\in C^{2-}([0,1])$. Moreover, 
\begin{align}\label{barq1}
\barq(p) &=\beta+(1+\theta)\int_{0}^{1}\barq(t) \inw'(t)\dt-(u')^{-1}\big(\opl'(p)\big),
\end{align}
and thus 
\begin{align*}
\barq'(p) &=\frac{\opl''(p)}{-u''\big((u')^{-1}\big(\opl'(p)\big)\big)},\quad\mbox{a.e.}\;p\in[0,1].
\end{align*}
Thanks to \citelem{op2}, $\barq$ satisfies \eqref{vi001}, so
\begin{multline*} 
\min\Bigg\{\max\bigg\{\frac{\opl''(p)}{-u''\big((u')^{-1}\big(\opl'(p)\big)\big)}-\hbar(p), \;\opl(0) (1+\theta)(1-\inw(p))-\opl(p)\bigg\},\\
\hfill\;\frac{\opl''(p)}{-u''\big((u')^{-1}\big(\opl'(p)\big)\big)}\Bigg\}=0,\quad\mbox{a.e.}\;p\in[0,1].
\end{multline*} 
By virtue of \citelem{tech1}, the above equation is equivalent to the ODE in \eqref{vi002}. We now show the last equation in \eqref{vi002}. Multiplying $\inw'(p)$ on both sides in \eqref{barq1} and then integrating on $[0,1]$, we get
\begin{align*} 
\int_{0}^{1}\barq(p)\inw'(p)\ddp &=\beta+(1+\theta)\int_{0}^{1}\barq(t) \inw'(t)\dt-\int_{0}^{1}(u')^{-1}\big(\opl'(p)\big)\inw'(p)\ddp,
\end{align*}
so 
\begin{align*} 
\beta &=\int_{0}^{1}(u')^{-1}\big(\opl'(p)\big)\inw'(p)\ddp-\theta\int_{0}^{1}\barq(t) \inw'(t)\dt.
\end{align*}
When $p=0$, \eqref{barq1} reduces to 
\begin{align*}
0&=\beta+(1+\theta)\int_{0}^{1}\barq(t) \inw'(t)\dt-(u')^{-1}\big(\opl'(0)\big).
\end{align*}
Canceling the terms $\int_{0}^{1}\barq(t) \inw'(t)\dt$ from the above two equations, we obtain the last equation in \eqref{vi002}. It is left to show that $\opl$ is a good function. Because $\barq$ is non-decreasing, by definition $\opl$ is convex and $\opl'>0$. Suppose $q_0=w(\ms)$. Since $\barq=0$ on $[0,q_0]$, by definition $\opl'$ is a constant on $[0,q_0]$. For $p\in(q_0,1)$ such that $\opl(0) (1+\theta)(1-\ms)-\opl(p)>0$, we have $\opl(0) (1+\theta)(1-\inw(p))\geq \opl(0) (1+\theta)(1-\ms)>\opl(p)$, which together with \eqref{vi002} implies $\opl''(p)=0$ a.e.. Since $\opl'$ is continuous, we conclude $\opl'$ is a constant on the set $\{p\in[0,1]: \opl(0) (1+\theta)(1-\ms)-\opl(p)\geq 0\}$. Therefore, $\opl$ is a good function. 

\item 
Now suppose $\opl\in C^{2-}([0,1])$ is a solution to \eqref{vi002}. We set 
\begin{align*}
\barq(p) &=(u')^{-1}\big(\opl'(0)\big)-(u')^{-1}\big(\opl'(p)\big).
\end{align*}
Then $\barq$ is absolutely continuous, $\barq(0)=0$ and 
\begin{align*}
\barq'(p)=\frac{\opl''(p)}{-u''\big((u')^{-1}\big(\opl'(p)\big)\big)}, \quad\mbox{a.e.}\;p\in[0,1].
\end{align*} 
By virtue of \citelem{op2}, we can rewrite the ODE in \eqref{vi002} as 
\begin{multline*} 
\min\Bigg\{\max\bigg\{\frac{\opl''(p)}{-u''\big((u')^{-1}\big(\opl'(p)\big)\big)}-\hbar(p), \;\opl(0) (1+\theta)(1-\inw(p))-\opl(p)\bigg\},\quad\\
\;\frac{\opl''(p)}{-u''\big((u')^{-1}\big(\opl'(p)\big)\big)}\Bigg\}=0, \quad\mbox{a.e.}\;p\in[0,1],
\end{multline*} 
that is, 
\begin{multline*} 
\min\bigg\{\max\Big\{ \barq'(p)-\hbar(p), \;\opl(0) (1+\theta)(1-\inw(p))-\opl(p)\Big\}, \;\barq'(p)\bigg\}=0, \quad\mbox{a.e.}\;p\in[0,1].
\end{multline*} 
In other words, $\barq$ satisfies the ODE in \eqref{vi001}. 
We now show \eqref{opldef}. The last equation in \eqref{vi002} implies 
\begin{align*}
(u')^{-1}\big(\opl'(0)\big)
&=\beta+(1+\theta)\int_{0}^{1}\Big[ (u')^{-1}\big(\opl'(0)\big)-(u')^{-1}\big(\opl'(t)\big) \Big]\inw'(t)\dt,
\end{align*}
whose right hand side by definition is equal to
\begin{align*}
\beta+(1+\theta)\int_{0}^{1}\barq(t) \inw'(t)\dt.
\end{align*}
Hence 
\begin{align*}
\barq(p) &=(u')^{-1}\big(\opl'(0)\big)-(u')^{-1}\big(\opl'(p)\big)\\
&=\beta+(1+\theta)\int_{0}^{1}\barq(t) \inw'(t)\dt-(u')^{-1}\big(\opl'(p)\big),
\end{align*}
which implies $\beta+(1+\theta)\int_{0}^{1}\barq(t) \inw'(t)\dt-\barq(p)=(u')^{-1}\big(\opl'(p)\big)\geq 0$ so that 
\eqref{bigbeta} is satisfied and 
\begin{align*}
\opl'(p)&=u'\Big(\beta+(1+\theta)\int_{0}^{1}\barq(t) \inw'(t)\dt-\barq(p)\Big).
\end{align*}
Thanks to the boundary condition $\opl(1)=0$ in \eqref{vi002}, the above equation implies \eqref{opldef}. By \citelem{op2}, we see $\barq$ is the optimal solution to problem \eqref{opi3}. Consequently, by \citelem{optimals1}, $\overline{R}$ is the optimal solution to problem \eqref{opi}. 
\end{enumerate}
The above argument also shows that any solution $\opl$ in $C^{2-}([0,1])$ to \eqref{vi002} should satisfy \eqref{opldef}. Since the optimal solution $\barq$ to problem \eqref{opi3} is unique, we conclude that \eqref{vi002} has exactly one solution in $C^{2-}([0,1])$. By the first assertion, the solution is a good function. 
\end{proof} 
\par
As \eqref{vi001} is an OIDE while \eqref{vi002} is an ODE, intuitively speaking, the latter is easier to study than the former. 
\par
We have never seen an ODE like \eqref{vi002} in the financial economics literature. At first sight it looks like a standard double-obstacle problem, however, the obstacles are put on the highest (second) order gradient of the unknown function; by contrast, they are usually put on the lower order gradient(s) in the literature (see, e.g., Dai and Yi \cite{DY06}, Dai, Xu and Zhou \cite{DXZ10} with obstacles on the first order gradients). To the best of our knowledge, it is the first time that such type of double-obstacle problems appears in the financial economics literature. See \citeremark{studyODE} below for further discussion on it.

\begin{remark}
By \citeremark{G=01}, the function $\opl$ defined in \eqref{opldef3} satisfies 
\begin{align*} 
\opl(p)=u'\Big(\beta+(1+\theta)\int_{0}^{1}\barq(t) \inw'(t)\dt\Big)(p-\ms)+\opl(\ms)
\end{align*} 
for $p\leq w(\ms)$. 
Using the fact that $\hbar(p)=0$ for $p\leq w(\ms)$, this can also be deduced from \eqref{vi002} directly. 
\end{remark}

\begin{remark}\label{specialcases}
We may simplify \eqref{vi002} for the most widely used three utility functions.
\begin{itemize}
\item For the exponential utility $u(x)=-\alpha^{-1}e^{-\alpha x}$, $\alpha>0$, \eqref{vi002} reduces to 
\begin{equation} \label{vi002-1}
\begin{cases}
\min\Big\{\max\big\{ \opl''(p)-\alpha\hbar(p)\opl'(p), \;\opl(0) (1+\theta)(1-\inw(p))-\opl(p)\big\},\quad\\
\hfill \opl''(p)\Big\}=0, \quad\mbox{a.e.}\;p\in[0,1], \\
\opl(1)=0, \quad
\beta=\alpha^{-1}\Big( (\log \opl'(t) ) \theta-(1+\theta)\int_{0}^{1} (\log \opl'(t) )\inw'(t)\dt\Big).
\end{cases}
\end{equation} 
In this case the operator grows linearly in $\opl'$. 
\item For the power utility $u(x)=\alpha^{-1} x^{\alpha}$, $\alpha<0$ or $0<\alpha<1$, \eqref{vi002} reduces to 
\begin{equation} \label{vi002-2}
\begin{cases}
\min\Big\{\max\big\{ \opl''(p)-(1-\alpha)\hbar(p) (\opl'(p))^{\frac{2-\alpha}{1-\alpha}}, \;\opl(0) (1+\theta)(1-\inw(p))-\opl(p)\big\},\quad\\
\hfill \opl''(p)\Big\}=0, \quad\mbox{a.e.}\;p\in[0,1], \\
\opl(1)=0, \quad
\beta=(1+\theta)\int_{0}^{1} (\opl'(t) )^{\frac{1}{\alpha-1}}\inw'(t)\dt-(\opl'(0) )^{\frac{1}{\alpha-1}}\theta.
\end{cases}
\end{equation} 
If $\alpha<0$, then $1<\frac{2-\alpha}{1-\alpha}<2$, so the operator is between linear and quadratic growth in $\opl'$. If $0<\alpha<1$, then $\frac{2-\alpha}{1-\alpha}>2$, so the operator is beyond quadratic growth in $\opl'$, making the problem challenging. 

\item For the logarithmic utility $u(x)=\log x$, \eqref{vi002} reduces to 
\begin{equation} \label{vi002-3}
\begin{cases}
\min\Big\{\max\big\{ \opl''(p)-\hbar(p) (\opl'(p))^{2}, \;\opl(0) (1+\theta)(1-\inw(p))-\opl(p)\big\},\quad\\
\hfill \opl''(p)\Big\}=0, \quad\mbox{a.e.}\;p\in[0,1], \\
\opl(1)=0, \quad
\beta=(1+\theta)\int_{0}^{1} (\opl'(t))^{-1}\inw'(t)\dt-(\opl'(0))^{-1} \theta.
\end{cases}
\end{equation} 
This can be regarded as the limit case $\alpha=0$ in \eqref{vi002-2}.
\end{itemize} 
\end{remark}
\par
Although \eqref{vi002} is an ODE, it is not easy to study for the following reasons: First, $\opl(0)$ appears in the operator, so the operator is nonlocal, which is in general harder than problems with local operator. Second, as mentioned earlier, the obstacles are put on the highest order gradient of the unknown function rather than lower order gradients; the corresponding penalty approximating problem is a fully nonlinear one rather than a semi-linear one. Third, the operator $\opl''+\hbar u''\big((u')^{-1}(\opl')\big)$ may not be linear growth in $\opl'$ (see, e.g,., \eqref{vi002-2} and \eqref{vi002-3}). Fourth, the last boundary condition is nonlocal and not the classical mixed Dirichlet and Neumann boundary conditions. We cannot find any existing numerical method that can be applied directly to solve \eqref{vi002}. In fact, its solvability including the existence and uniqueness of the solution is an issue from the pure ODE point of view, although we have resolved this issue from the pure optimization point of view. See \citeremark{studyODE} below for further discussion.

\begin{remark}
The author of this paper introduced a change of variables and relaxation method in \cite{X16} to solve the first-type quantile optimization problem (e.g. the problem in \cite{XZ16}). The optimal solution is expressed via the concave envelope of some known function (we remark that a concave envelope can be expressed as the solution to a single-obstacle problem). The model does not involve the compatibility constraint, and it can be regarded as the special case $\hbar(p)=+\infty$ for $p\in (w(\ms),1]$ in our model. In this case, 
\[ \opl''(p)+\hbar(p)u''\big((u')^{-1}\big(\opl'(p)\big)\big)=-\infty, \quad p\in (w(\ms),1],\]
so the ODE in \eqref{vi002} reduces to 
\begin{equation*} 
\begin{cases}
\opl''(p)=0, & \quad \mbox{a.e.}\;p\in[0, w(\ms)], \\
\min\Big\{\opl(0) (1+\theta)(1-\inw(p))-\opl(p),\; \opl''(p)\Big\}=0, & \quad\mbox{a.e.}\;p\in(w(\ms),1].
\end{cases}
\end{equation*}From this, one can reveal the optimal solution for \cite{XZ16} (as well as \cite{X16}). Roughly speaking, the first-type quantile optimization problems only involve one-side constraint, while the second-type problems require two-side constraint, so the classical single-obstacle ODE becomes a double-obstacle ODE. As a result, no closed-form solution is available for the second-type problem in general. 
\end{remark}

\par
It is very hard to find a numerical scheme to solve \eqref{vi002} directly due to the last nonlocal condition. Recalling that our original target is to find all the PO contracts, this inspires us to consider all the possible values of $\beta$ (as a set) rather than a particular one. This allows us forgetting the last nonlocal condition in \eqref{vi002} so that it can be solved. The following result will play a key role in designing a numerical scheme to calculate all the PO contracts in the next section. 
\begin{coro}\label{optimalall}
A function $\opl$ in $C^{2-}([0,1])$ corresponds to a PO contract if and only if it satisfies 
\begin{equation} \label{vi003}
\begin{cases}
\min\Big\{\max\big\{ \opl''(p)+\hbar(p)u''\big((u')^{-1}\big(\opl'(p)\big)\big), \;\inone (1+\theta)(1-\inw(p))-\opl(p)\big\},\quad\\
\hfill \opl''(p)\Big\}=0, \quad\mbox{a.e.}\;p\in[0,1], \\
\opl(0)=\inone, \quad\opl(1)=0,
\end{cases}
\end{equation} 
for some $\inone<0$; 
in which case
\[\barq(p):=(u')^{-1}\big(\opl'(0)\big)-(u')^{-1}\big(\opl'(p)\big)\] and 
\[\overline{R}(x):=(u')^{-1}\big(\opl'(0)\big)-(u')^{-1}\big(\opl'\big(w(F_X(x))\big)\big)\]
are, respectively, the optimal solutions to problems \eqref{opi3} and \eqref{opi} with
\[\beta:=(1+\theta)\int_{0}^{1} (u')^{-1}\big(\opl'(t)\big)\inw'(t)\dt-(u')^{-1}\big(\opl'(0)\big) \theta.\]
Moreover, the solution $\opl$ to \eqref{vi003} is a good function. 
\end{coro}

\begin{proof}
This is an immediate consequence of \citethm{main1}. We leave the details to the interested readers. 
\end{proof}
\par
The ODE \eqref{vi003} is a problem with Dirichlet boundary conditions, but we failed to find an existing numerical scheme that can be applied directly to solve it. Therefore, we will provide an alternative way to solve it in the following section.

\subsection{A numerical scheme to solve the problem}\label{numerical}
By \citethm{main1}, solving problems \eqref{opi3} and \eqref{opi} reduces to solving \eqref{vi002}. But for a given $\beta$, it is very hard to directly solve \eqref{vi002}, even numerically. \citecoro{optimalall} motives us to solve the problem for all the possible values of $\beta$. Hence we propose the following numerical scheme. It turns out that we can use this method to solve \eqref{vi002} and \eqref{opi} for any fixed $\beta$ as well. 
\begin{enumerate}
\item First fix any value $\inone<0$. Then, for each $\inprime>0$, consider the following problem\footnote{Clearly the solution $\opl$ to \eqref{vi002} must be convex and satisfy $\opl(1)=0$ and $\opl'>0$ by \citethm{main1}. Therefore, it suffices to consider $\inprime> 0$ and $\inone<0$.}
\begin{equation} \label{vi004}
\begin{cases}
\min\Big\{\max\big\{ \opl''(p)+\hbar(p)u''\big((u')^{-1}\big(\opl'(p)\big)\big), \;\inone (1+\theta)(1-\inw(p))-\opl(p)\big\},\quad\\
\hfill\opl''(p)\Big\}=0, \quad\mbox{a.e.}\;p\in[0,1], \\
\opl(0)=\inone, \quad \opl'(0)=\inprime.
\end{cases}
\end{equation} 
This is an initial value double-obstacle problem with a semi-linear \emph{local} operator. We can solve it, say, by finite difference method. We denote its solution by $\opl_{\inone,\inprime}$. 
\item By comparison theorem for nonlinear ODE (see, e.g. Lieberman \cite{Li96}), we can show that $\opl_{\inone,\inprime}$ is non-decreasing in $\inprime$ and there exists a ${\inprime}({\inone})$ (depending on $\inone$) such that $\opl_{\inone, {\inprime}({\inone})}(1)=0$ (assuming $\hbar$ is a continuous function). Moreover, the map $\inone\mapsto {\inprime}({\inone})$ is non-increasing. By virtue of this, one can check that $\opl_{\inone, {\inprime}({\inone})}$ solves \eqref{vi002} with $\beta$ replaced by 
\begin{align} \label{betain}
{\beta}({\inone})=(1+\theta)\int_{0}^{1} (u')^{-1}\big(\opl_{\inone, {\inprime}({\inone})}'(t)\big)\inw'(t)\dt- (u')^{-1} ({\inprime}({\inone}))\theta.
\end{align} 
Since $\opl_{\inone,\inprime}$ is convex, we have $\opl_{\inone,\inprime}(p)\geq \opl_{\inone,\inprime}(0)+\opl'_{\inone,\inprime}(0)p$ for $p\in[0,1]$. In particular, $0=\opl_{\inone, {\inprime}({\inone})}(1)\geq \opl_{\inone, {\inprime}({\inone})}(0)+\opl_{\inone, {\inprime}({\inone})}'(0)=\inone+\inprime({\inone})$, that is $\inprime({\inone})\leq -\inone.$ Therefore, it suffices to consider $0<\inprime\leq -\inone$ in the first step in order to find $\inprime({\inone})$.
\item The ODE \eqref{vi002} has at most one solution for each fixed $\beta$, so different values of $\inone$ must lead to different values of ${\beta}({\inone})$. Hence we will get all the PO contracts if we go through all $\inone<0$. Since $\opl'_{\inone,\inprime}$ is continuous in $\inone$ and $\inprime$, ${\beta}({\inone})$ is continuous with respective to $\inone$. The map $\inone\mapsto {\beta}({\inone})$ is injective and continuous, so ${\beta}({\inone})$ is a monotone function of $\inone$. 
The solution to \eqref{vi004} is evidently convex, so by \eqref{betain} and the concavity of $u$, one has 
\begin{align*} 
{\beta}({\inone})\leq (1+\theta)\int_{0}^{1} (u')^{-1}\big(\opl_{\inone, {\inprime}({\inone})}'(0)\big)\inw'(t)\dt- (u')^{-1} ({\inprime}({\inone}))\theta=(u')^{-1} ({\inprime}({\inone})).
\end{align*} 
This implies \[\lim_{\inone\to -\infty}{\beta}({\inone})=\lim_{{\inprime}({\inone})\to \infty}{\beta}({\inone})\leq 
\lim_{{\inprime}({\inone})\to \infty}(u')^{-1} ({\inprime}({\inone}))<\infty.\] We conclude that ${\beta}({\inone})$ must be strictly increasing in $\inone$ since there is no upper bound for $\beta$ in problem \eqref{opi}. Because ${\beta}({\inone})$ is monotone, we can solve \eqref{vi002} numerically for any feasible $\beta$, say, by bisection method. 
\end{enumerate}

\begin{remark}\label{studyODE}
The ODEs \eqref{vi002}, \eqref{vi003} and \eqref{vi004} are interesting in their own right in ODE theory and detailed study of them requires delicate analysis that is beyond the scope of this paper, we leave the details in a forthcoming paper. Here we just briefly introduce the main idea to tackle them. To study problem \eqref{vi004}, for instance, it suffices to study the following ODE for $(\opl,\opp)$: 
\begin{equation} \label{vi002p1}
\begin{cases}
\min\Big\{\opp(p),\;\opl''(p)\Big\}=0, \\
\opp(p)=\max\Big\{ \opl''(p)+\hbar(p)u''\big((u')^{-1}\big(\opl'(p)\big)\big), \;\inone (1+\theta)(1-\inw(p))-\opl(p)\Big\},\\
\opl(0)=\inone, \quad \opl'(0)=\inprime.
\end{cases}
\end{equation} 
We first introduce the following penalty approximation ODE for the above one: 
\begin{equation} \label{vi002p2}
\begin{cases} 
\opl''(p)+\penal_{1,\ep} (\opp(p))=0,\\
\opl''(p)+\hbar(p)u''\big((u')^{-1}\big(\opl'(p)\big)\big)-\opp(p)- \penal_{2,\ep}\big(\opp(p)+\opl(p)-\inone(1+\theta)(1-\inw(p))\big)=0,\\
\opl(0)=\inone, \quad \opl'(0)=\inprime,
\end{cases}
\end{equation}
where $\penal_{1,\ep}$ and $\penal_{2,\ep}$ are some proper penalizing functions. The ODE \eqref{vi002p2} is an initial value problem with semi-linear operator and we can show that it has a solution for each $\ep>0$ under certain reasonable assumptions on its coefficients. Finally we show the solution convergences to a solution to problem \eqref{vi002p1} as $\ep\to0$. The properties (such as monotonicity in $\inone$ and $\inprime$) of the solution to \eqref{vi002p1} inherit that of \eqref{vi002p2}. 
\end{remark}

\subsection{A numerical example} \label{numericalexample}
In this section we perform a numerical example to illustrate the theoretical results obtained thus far. 

The function $\inw$ is set as 
\begin{align*}
\inw(p)=\begin{cases}
\dsp\frac{25}{12}(p-2p^2), &\quad p\in[0, \frac{1}{5}];\bigskip\\
\dsp\frac{1}{444}\big(625 p^3-375 p^2+260 p+69\big), &\quad p\in(\frac{1}{5}, \frac{2}{5}];\bigskip\\
\dsp\frac{1}{444}\big(375 p^2-40 p+109\big), &\quad p\in(\frac{2}{5},1], 
\end{cases}
\end{align*} 
The probability weighting function $w$ is defined as the inverse of $\inw$. We can see that both $w$ and $\inw$ are $C^{2-}([0,1])$ function with positive derivative functions. 
Their pictures are illustrated in \citefig{fig_w}.\smallskip
\begin{figure}[H] 
\center{ 
\includegraphics[width=0.47\textwidth]{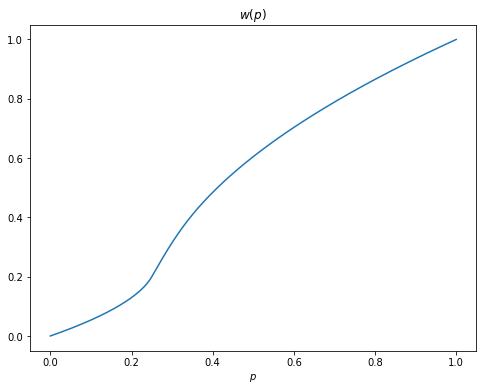}\qquad
\includegraphics[width=0.47\textwidth]{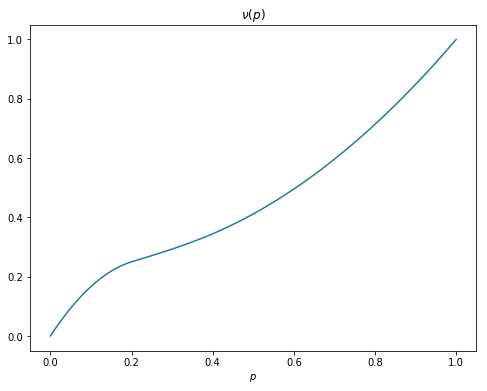}}\smallskip \caption{\label{fig_w} 
The probability weighting function $w$ and its inverse $\inw$.} 
\end{figure}
We choose a power utility function $u(x)=2\sqrt{x}$, set the relative safety loading $\theta=0.2$, and define the quantile of the potential loss $X$ by 
\begin{align*}
F_X^{-1}(\inw(p))=
\begin{cases}
0, &\quad p\in[0,\frac{1}{5}];\medskip\\
\dsp\frac{2}{3}(1875 p^4-2500 p^3+1200 p^2-240 p+617)\\
\dsp\qquad-\left(\frac{74}{375 p^2-150 p+ 52}\right)^{2}, &\quad p\in(\frac{1}{5}, \frac{2}{5}];\bigskip\\ 
\dsp\frac{14}{3}-\left(\frac{666}{-2500 p^3+4125 p^2-750 p+268}\right)^{2}, &\quad p\in(\frac{2}{5}, 1]. 
\end{cases}
\end{align*}
We can show $F_X^{-1}$ is continuous differentiable and fulfills \citeassmp{ass1} with a mass $\ms=\frac{1}{4}$ at 0. Moreover, the corresponding function $\hbar$ is a continuous function. 
The pictures of $F_X^{-1}$ and $\hbar$ are shown in \citefig{fig_f}.\smallskip
\begin{figure}[H] 
\center{ 
\includegraphics[width=0.46\textwidth]{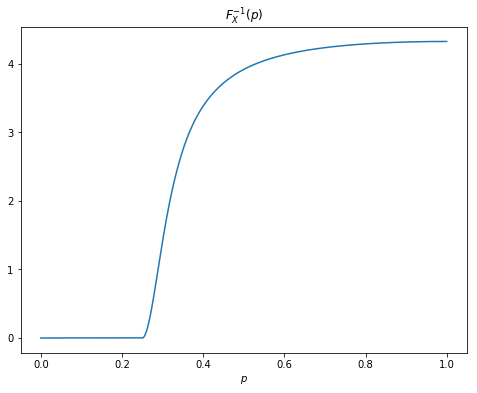}\qquad
\includegraphics[width=0.477\textwidth]{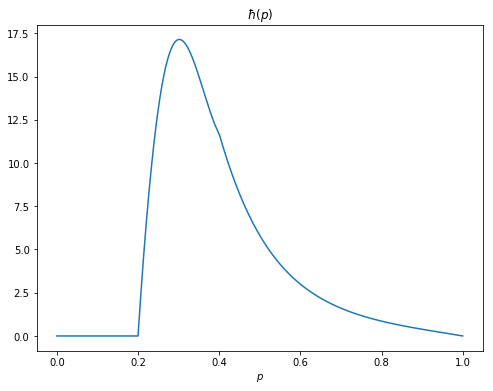}}\smallskip
\caption{\label{fig_f} The quantile function of $X$ and the function $\hbar$.} 
\end{figure} 
Under the above setting, we use the numerical scheme proposed in the previous section to compute the function $\opl$ and the optimal retention $\overline{R}$ for $\beta=-0.928$. Our scheme shows that $\inone=-1$ and $\inprime=0.5$ in this case. 
The pictures of $\opl$ and $\overline{R}$ are presented in \citefig{fig_phi} and \citefig{fig_r}, respectively. 
\begin{figure}[H] 
\center{ 
\includegraphics[width=0.6\textwidth]{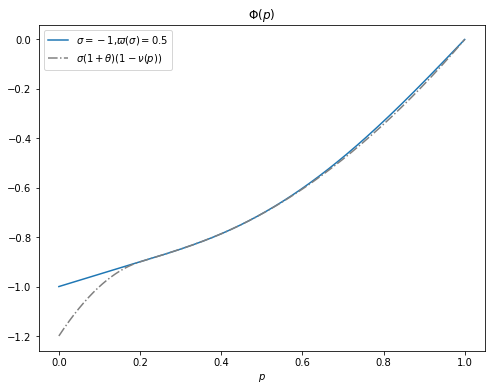}}
\caption{\label{fig_phi} The function $\opl$. } 
\end{figure} 
\begin{figure}[H] 
\center{ 
\includegraphics[width=0.6\textwidth]{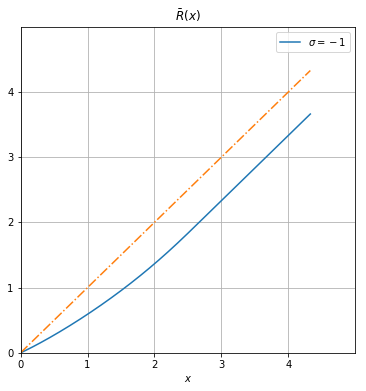}}
\caption{\label{fig_r} The optimal retention $\overline{R}$. } 
\end{figure} 
It can be seen from \citefig{fig_phi} that $\opl(p)$ is convex and different from $\inone(1+\theta)(1-\inw(p))$ when $p$ is near to 0 or 1. When $p$ is small, $\opl(p)$ is linear which is due to the fact that $\opl(p)$ is a good function. When $p$ is in the middle range, $\opl(p)$ coincides with $\inone(1+\theta)(1-\nu(p))$, which indicates that the derivative constraint is not likely tight, namely $0<\overline{R}'<1$. This indeed can be observed from \citefig{fig_r}. When $p$ is large, $\opl(p)$ dominates $\inone(1+\theta)(1-\nu(p))$, which corresponds to the tight case, namely $\overline{R}'=1$. This can also be seen from \citefig{fig_r}.

\section{On the PO contracts}\label{property}
In this section we discuss some properties of the PO contracts. 

\subsection{ PO contracts of deductible type}

In this section we investigate under what conditions a deductible compensation is PO. Our first result reveals the classical assertion that all the PO contracts are of deductible type in the EU theory framework (see, e.g. \cite{SZ71}). 
\begin{prop}\label{EUcase}
If there is no probability distortion, then all the PO contracts are of deductible type.
\end{prop}

\begin{proof}
Suppose there is no probability distortion, i.e., $w(p)\equiv p$. Then \eqref{vi003} reduces to 
\begin{equation} \label{vi005}
\begin{cases}
\min\Big\{\max\big\{ \opl''(p)+h(p)u''\big((u')^{-1}\big(\opl'(p)\big)\big), \;\inone (1+\theta)(1-p)-\opl(p)\big\},\quad\\
\hfill \opl''(p)\Big\}=0, \quad\mbox{a.e.}\;p\in[0,1], \\
\opl(0)=\inone<0, \quad\opl(1)=0.
\end{cases}
\end{equation} 
Since $\opl$ is continuous and convex, $f(p)=\inone (1+\theta)(1-p)-\opl(p)$ is a continuous concave function on $[0,1]$ with boundary values $f(0)=\inone\theta<0$ and $f(1)=0$. 
Let 
\[p_0=\sup\{p\in[0,1]\colon f(p)< 0\}.\]
Then $0<p_0\leq 1$, $f<0$ on $[0,p_0)$, $f(p_0)=0$ and $f\geq 0$ on $[p_0,1]$. By \eqref{vi005}, we obtain $\opl''(p)+h(p)u''\big((u')^{-1}\big(\opl'(p)\big)\big)=0$ for $\mbox{a.e.}\;p\in[0,p_0)$. As $f$ is concave, there are two possible cases on $[p_0,1]$: 
\begin{enumerate} 
\item Either $f=0$ on $[p_0, 1]$, which trivially implies $\opl''(p)=(\inone (1+\theta)(1-p)-f(p))''=0$ for $p\in(p_0,1)$.
\item Or $f>0$ on $(p_0,1)$, which by \eqref{vi005} also implies that $\opl''(p)=0$ for a.e. $p\in(p_0,1)$. 
\end{enumerate} 
Therefore, in both cases, we have
\begin{align*}
\barq'(p)=\frac{\opl''(p)}{-u''\big((u')^{-1}\big(\opl'(p)\big)\big)}=
\begin{cases}
h(p), &\quad \mbox{a.e.}\;p\in[0,p_0);\\
0, &\quad \mbox{a.e.}\;p\in[p_0,1].
\end{cases}
\end{align*}
By virtue of $\barq(0)=0$ and \eqref{def:h}, we obtain
\begin{align*}
\barq(p)=\int_0^{\min\{p,\;p_0\}}h(t)\dt=\min\big\{F_{X}^{-1}(p),\;F_{X}^{-1}(p_0)\big\}.
\end{align*}
Thanks to \eqref{optimals2}, we see the corresponding compensation 
\begin{align*}
\overline{I}(x)\equiv x-\overline{R}(x)\equiv x- \barq\big(F_X(x)\big)\equiv x-\min\{x,\;d\}=\max\{x-d, \;0\}
\end{align*}
is of deductible type, where $d=F_{X}^{-1}(p_0)$. In particular the compensation becomes the full coverage compensation if $p_0\leq \ms$.
\end{proof}

\begin{remark}\label{partialdistortion}
By \citeremark{G=0}, the optimal solution solution to \eqref{opi} does not depend on the shape of $w$ on $[0,\ms)$, so \citeprop{EUcase} still holds if the condition is replaced by $w(p)=p$ for $p\in[\ms,1]$.
\end{remark}

When the probability weighting function is nontrivial, the following result answers the question completely by providing an equivalent condition. 

\begin{thm}\label{special1}
The optimal solution to problem \eqref{opi} corresponds to a deductible compensation with deductible $d$ if and only if 
\begin{align}\label{cond1}
\begin{cases}
(1+\theta)(1-\inw(p))\int_{0}^{1}u'(g(s))\ds\geq \int_{p}^{1}u'(g(s))\ds,\quad p\in[0, w(F_X(d))];\\
(1+\theta)(1-\inw(p))\int_{0}^{1}u'(g(s))\ds\leq \int_{p}^{1}u'(g(s))\ds,\quad p\in [w(F_X(d)), 1],
\end{cases}
\end{align} 
where
\begin{align*}
g(s) &=\beta+(1+\theta)\BE{\min \{X,d\}}-\min\big\{F_{X}^{-1}(\inw(s)),\;d \big\}.
\end{align*} 
Moreover, the deductible $d$ satisfies $F_X(d)\geq \frac{\theta}{1+\theta}$. 
\end{thm}

\begin{proof}
By \eqref{optimals2}, $R(x)=\min\{x,\; d\}$ is the optimal solution to problem \eqref{opi} if and only if 
\[\barq(p):=\min\big\{F_{X}^{-1}(\inw(p)),\;d\big\}\]
is the optimal solution to problem \eqref{opi3}, which by \citethm{main1} is also equivalent to 
\begin{align*}
\baropl(p):=-\int_{p}^{1}u'\Big(\beta+(1+\theta)\int_{0}^{1}\barq(t) \inw'(t)\dt-\barq(s)\Big)\ds
\end{align*} 
is a solution to \eqref{vi002} in $C^{2-}([0,1])$. 
Thanks to \eqref{h2}, we have 
\begin{align*}
\frac{\baropl''(p)}{-u''\big((u')^{-1}\big(\baropl'(p)\big)\big)}=\barq'(p)=\begin{cases}
\hbar(p), &\mbox{a.e.}\;p\in[0, w(F_X(d))];\\
0, & \;p\in (w(F_X(d)),1].
\end{cases}
\end{align*} 
By virtue of this, one can check $\baropl$ is a solution to \eqref{vi002} holds true if and only if 
\begin{align*}
\begin{cases}
\baropl(0)(1+\theta)(1-\inw(p))-\baropl(p) \leq 0,\quad p\in[0, w(F_X(d))];\\
\baropl(0)(1+\theta)(1-\inw(p))-\baropl(p) \geq 0, \quad p\in[w(F_X(d)),1].
\end{cases}
\end{align*} 
But this is equivalent to \eqref{cond1} because 
\begin{align} \label{eq1}
\baropl(p)&=-\int_{p}^{1}u'\Big(\beta+(1+\theta)\int_{0}^{1}\min\big\{F_{X}^{-1}(\inw(t)),\;d\big\}\inw'(t)\dt-\min\big\{F_{X}^{-1}(\inw(s)),\;d\big\}\Big)\ds\nn\\
&=-\int_{p}^{1}u'\Big(\beta+(1+\theta)\int_{0}^{1}\min\big\{F_{X}^{-1}(t),\;d\big\}\dt-\min\big\{F_{X}^{-1}(\inw(s)),\;d\big\}\Big)\ds\nn\\
&=-\int_{p}^{1}u'\Big(\beta+(1+\theta)\BE{\min \{X,d\}}-\min\big\{F_{X}^{-1}(\inw(s)),\;d\big\}\Big)\ds\nn\\
&=-\int_{p}^{1}u'(g(s))\ds.
\end{align} 
When $p=w(F_X(d))$, the condition \eqref{cond1} implies 
\[(1+\theta)(1-F_X(d))=\frac{ \int_{w(F_X(d))}^{1}u'(g(s))\ds}{\int_{0}^{1}u'(g(s))\ds}.\]
The right hand side is clearly no more than 1, so $F_X(d)\geq \frac{\theta}{1+\theta}$. 
\end{proof}

\begin{coro}\label{coro1}
If $\theta>\frac{\ms}{1-\ms}$, then the full coverage compensation is not optimal for any RDU insured. 
\end{coro}

\begin{proof}
By \citethm{special1}, we should have $F_X(d)\geq \frac{\theta}{1+\theta}>\ms$, so $d>0$.
\end{proof}
In particular, when $X$ has no mass at $0$, namely $\ms=0$, the condition in \citecoro{coro1} is satisfied, therefore the full coverage compensation is not optimal for any RDU insured. On the other hand, by the proof of \citeprop{EUcase} and \citeremark{partialdistortion}, the full coverage compensation can be PO at least when $ \ms>0$ and $w(p)=p$ for $p\in[ \ms,1]$. Therefore, the relative safety loading and the mass of $X$ at 0 play the critical role in determining the type of the PO contracts. Indeed, when the relative safety loading is small or $X$ has a large mass at 0, namely $0< \theta<\frac{\ms}{1-\ms}$, the full coverage compensation can also be PO even if the probability weighting function is nontrivial. This will be shown in the following section. 

\begin{remark}
When $w(F_X(d))\leq p\leq 1$, we have 
\begin{align*} 
\int_{p}^{1}u'(g(s))\ds &=\int_{p}^{1}u'\Big(\beta+(1+\theta)\BE{\min \{X,d\}}-\min\big\{F_{X}^{-1}(\inw(s)),\;d\big\}\Big)\ds\nn\\
&=\int_{p}^{1}u'\big(\beta+(1+\theta)\BE{\min \{X,d\}}-d\big)\ds\nn\\
&=u'\big(\beta+(1+\theta)\BE{\min \{X-d,0\}}\big)(1-p).
\end{align*} 
\end{remark} 

\begin{remark}
Clearly one can use our idea to find equivalent conditions for the optimal compensations being a special form such as the proportional coverage compensations. We leave this to the interested readers. 
\end{remark}

\begin{remark}
Xu, Zhou and Zhuang \cite{XZZ19} studied an optimal compensation problem, which is slight different from ours, but also under the RDU theory. Under certain assumptions (see \citeremark{remarkxzz}), they found that optimal compensation is of three-fold, that is 
\begin{align*}
I(x)=\max\big\{\min\{x,\;d_1\},\; x-d_2\big\}.
\end{align*}
Their conditions are sufficient but not necessary, by our method, one can find the equivalent conditions for the above compensation being optimal to their problem. 
\end{remark}

When \eqref{cond1} is not satisfied, a PO contract may not be of deductible type. We are interested in which situation a moral-hazard-free contract is PO, that is, it is optimal for at least one RDU insured; in other words, economically speaking, when such a contract is acceptable in insurance market. This will be addressed in the following section.

\subsection{When is a moral-hazard-free contract PO?}

In this section, we consider the following reverse problem. For a given potential loss $X$, a moral-hazard-free retention $R\in\setr$ and a relative safety loading $\theta>0$, is it possible to find at least one RDU insured such that the contract is optimal for her? Every RDU insured is characterized by her utility function $u$, probability weighting function $w$ and wealth level $\beta$. So the question can be mathematically stated as follows: for any given reasonable $(F_{X}, R, \theta)$, can we find a triple $(u, w, \beta)$ such that $R$ is optimal to problem \eqref{opi}? 

When the relative safety loading is big or $X$ has a small mass at 0, namely $\theta>\frac{\ms}{1-\ms}$, the answer is negative as we showed in \citecoro{coro1} that it is impossible to find a RDU insured such that the full coverage contract is optimal for her. By contrast, when the relative safety loading is small or $X$ has a large mass at 0, namely $0<\theta<\frac{\ms}{1-\ms}$, the following result gives an affirmative answer to the question. 
\begin{thm}\label{main2}
Suppose $R\in\setr$ is a moral-hazard-free retention, $X$ satisfies \citeassmp{ass1} and $0< \theta<\frac{\ms}{1-\ms}$. 
Then $R$ is optimal to problem \eqref{opi} for infinite many RDU insureds. 
\end{thm}
\begin{proof} 
By \citecoro{optimalall}, we only need to show that, for each good function $\opl$, there exist
infinite pairs $(u, w)$ such that $\opl$ is a solution to \eqref{vi003}, where $\inone=\opl(0)$, $\hbar(p)=\big(F_{X}^{-1}(\inw(p))\big)'$ and $\inw$ is the inverse function of $w$. 
\par
Define \[q_0=\inf\{p\in[0,1]: \opl(0) (1+\theta)(1-\ms)-\opl(p)<0\}.\]
Since $\opl(0)<\opl(1)=0$ and $0<(1+\theta)(1-\ms)<1$, by continuity, we have $0<q_0<1$ and $\opl(0) (1+\theta)(1-\ms)=\opl(q_0)$. 
Set any feasible $\inw$ on $[0,q_0]$ such that $\inw(q_0)=\ms$. 
By the definition of good function, we see $\opl$ is linear on $[0,q_0]$. Since $\inw(p)\leq \inw(q_0)=\ms$ for $p\in[0, q_0)$, $F_{X}^{-1}(\inw(p))=0$. Thus $\hbar(p)=\big(F_{X}^{-1}(\inw(p))\big)'=0$ and $\opl''(p)=0$ for $p\in[0, q_0)$. By virtue of this, 
\begin{multline*} 
\min\Big\{\max\big\{ \opl''(p)+\hbar(p)u''\big((u')^{-1}\big(\opl'(p)\big)\big), \;\opl(0) (1+\theta)(1-\inw(p))-\opl(p)\big\}, \;\opl''(p)\Big\}\qquad\quad\\
=\min\Big\{\max\big\{0, \;\opl(0) (1+\theta)(1-\inw(p))-\opl(p)\big\}, \;0\Big\}=0, \quad\mbox{a.e.}\;p\in[0,q_0].
\end{multline*} 
So the ODE in \eqref{vi003} is satisfied for a.e. $p\in[0, q_0]$. 
\par
We now consider the problem on $(q_0,1]$.
Let \[\inw(p)=1-\tfrac{\opl(p)}{\opl(0) (1+\theta)},\quad p\in(q_0,1].\] 
Since $\opl'>0$ and $\opl(0)<0$, $\inw$ is absolutely continuous and strictly increasing. Moreover, 
\begin{align*}
\inw(1)=1, \quad \inw(q_0+)=1-\tfrac{\opl(q_0+)}{\opl(0) (1+\theta)}=\ms=\inw(q_0),
\end{align*} 
so $\inw\in\setw$.
Because $\opl'(p)$ is continuous and non-decreasing and $F_X^{-1}(\inw(p))$ is strictly increasing on $[q_0,1]$, there exists a utility function $u$ which is in $C^2([0,\infty))$ with $u'>0$ and $u''<0$ such that 
\begin{align} \label{monotoneu}
\mbox{the mapping $\; p\mapsto (u')^{-1}(\opl'(p))+F_X^{-1}(\inw(p))$ is strictly increasing on $[q_0,1]$. }
\end{align} 
It then follows 
\[\frac{\opl''(p)}{-u''\big((u')^{-1}\big(\opl'(p)\big)\big)}-\big(F_{X}^{-1}(\inw(p))\big)'
=\big(-(u')^{-1}(\opl'(p))-F_X^{-1}(\inw(p))\big)'\leq 0, \quad\mbox{a.e.}\;p\in(q_0,1].\]
Notice 
$\opl(0) (1+\theta)(1-\inw(p))=\opl(p)$ for $p\in (q_0,1]$, so
\begin{multline*} 
\max\bigg\{\frac{\opl''(p)}{-u''\big((u')^{-1}\big(\opl'(p)\big)\big)}-\big(F_{X}^{-1}(\inw(p))\big)', \;\opl(0) (1+\theta)(1-\inw(p))-\opl(p)\bigg\}=0,\quad\mbox{a.e.}\; p\in(q_0,1]. 
\end{multline*} 
Because $\opl''\geq 0$, the above equation implies the ODE in \eqref{vi003} is satisfied on $(q_0,1]$. Since there are infinite many choices of $u$ that satisfies \eqref{monotoneu}, the claim is proved. 
\end{proof}

\begin{remark}
Once the utility function $u$ and the probability weighting function $w$ are chosen, one can determine the insured's $\beta$ by the last equation in \eqref{vi002}. Then $R$ will be optimal to problem \eqref{opi} for the insured $(u,w,\beta)$. 
\end{remark}

\begin{remark}
The marginal case $\theta=\frac{\ms}{1-\ms}$ requires delicate analysis, we leave it to the interested readers. 
\end{remark}
%

\section{ PO contracts with non-monotone compensations or retentions}\label{nonmonotone}
The {incentive compatibility} constraint is taken a priori into consideration in problem \eqref{opi0}, so the PO contracts are free of \emph{ex ante} moral hazard. One nature question is: what will the PO contracts look like if the {incentive compatibility} constraint is ignored in the problem formulation at beginning? We give complete answers to this question when the retentions or compensations are not necessary to be monotone. Since the analysis is similar to the case with incentive compatibility constraint, we just point out the main differences in the arguments and leave the details to the interest readers.

\subsection{PO contracts with non-monotone compensations}
If we only require the retentions to be monotone but put no constraint on the compensations,
that is, 
\begin{align} \label{compensation2}
\setr&=\Big\{R:[0,\infty)\to [0,\infty)\; \big|\; \mbox{$R$ is absolutely }\nn\\
&\qquad\quad~\mbox{ continuous with $R(0)=0$ and $ R'\geq 0$ a.e.}\Big\}.
\end{align} 
Then the corresponding set of compensations becomes 
\begin{align*} 
\setc&=\Big\{I:[0,\infty)\to [0,\infty)\; \big|\; \mbox{$I$ is absolutely }\nn\\
&\qquad\quad~\mbox{ continuous with $I(0)=0$ and $I'\leq 1$ a.e..}\Big\}.
\end{align*} 
In this case the compensations in $\setc$ may not be monotone. 

\begin{thm}\label{main3}
If the set of retentions $\setr$ in problem \eqref{opi} is replaced by \eqref{compensation2}, then a retention $\overline{R}\in\setr$ is optimal to problem \eqref{opi} if and only if it can be written as 
\begin{align} \label{optimalretention}
\overline{R}(x)=(u')^{-1}\big(\kappa\opll'(0)\big)-(u')^{-1}\big(\kappa\opll'\big(w(F_X(x))\big)\big),
\end{align} 
where $\kappa$ is the unique positive constant such that
\[\beta=(1+\theta)\int_{0}^{1} (u')^{-1}\big(\kappa\opll'(t)\big)\inw'(t)\dt-(u')^{-1}\big(\kappa\opll'(0)\big) \theta,\]
and $\opll$ is the convex envelope of $(1+\theta)(\inw(p)-1)$ on $[w(\ms),1]$ with boundary values $\opll(1)=0$ and $\opll(w(\ms))=(1+\theta)(\ms-1)$, and $\opll(p)=(p-w(\ms))\opll'(w(\ms))+(1+\theta)(\ms-1)$ for $p\in[0, w(\ms))$.
\end{thm}
\begin{proof}
The argument remains unchanged until \eqref{optimalcondition1-2}. For any $Q\in\setq$, since there is no upper bound limitation for $R'$ in \eqref{compensation2}, so is $Q'(p)$ for $p \in(w(\ms),1]$. 
Recall $Q=0$ on $[0,w(\ms)]$, so the optimality condition \eqref{optimalcondition1-2}
implies $\opl(p)\leq \opl(0) (1+\theta)(1-\inw(p))$ and \begin{align*}
\barq'(p)=\begin{cases}
\in[0,\infty), &\quad\text{if } \opl(p)=\opl(0) (1+\theta)(1-\inw(p));\\
0, &\quad\text{if } \opl(p)<\opl(0) (1+\theta)(1-\inw(p)),
\end{cases}\; \mbox{ for a.e. $p\in(w(\ms),1]$,}
\end{align*} 
which can be equivalently expressed as 
\begin{align*}
\min\Big\{\opl(0) (1+\theta)(1-\inw(p))-\opl(p),\;\barq'(p)\Big\}=0, \quad\mbox{a.e.}\;p\in(w(\ms),1].
\end{align*} 
This leads to the following analog of \eqref{vi002}, 
\begin{equation*} 
\begin{cases}
\opl''(p)=0, \hspace{200pt}\mbox{a.e.}\;p\in[0, w(\ms)], \\
\min\Big\{\opl(0) (1+\theta)(1-\inw(p))-\opl(p), \opl''(p)\Big\}=0, \quad\mbox{a.e.}\;p\in(w(\ms),1], \\
\opl(1)=0,\quad
\beta=(1+\theta)\int_{0}^{1} (u')^{-1}\big(\opl'(t)\big)\inw'(t)\dt-(u')^{-1}\big(\opl'(0)\big) \theta.
\end{cases}
\end{equation*} 
By virtue of \citelem{tech1} and $\opl(0)<0$, the ODE in above can be rewritten as 
\begin{equation*} 
\begin{cases}
\left(\tfrac{\opl(p)}{-\opl(0)}\right)''=0, &\quad\mbox{a.e.}\;p\in[0, w(\ms)], \\
\min\Big\{(1+\theta)(\inw(p)-1)-\tfrac{\opl(p)}{-\opl(0)},\; \left(\tfrac{\opl(p)}{-\opl(0)}\right)''\Big\}=0, &\quad\mbox{a.e.}\;p\in (w(\ms),1],\\
\frac{\opl(1)}{-\opl(0)}=0.
\end{cases}
\end{equation*} 
It is easy to check $\opll\in C^{2-}([0,1])$ and $\opll$ also satisfies the above ODE in place of $\frac{\opl(p)}{-\opl(0)}$.  
So $\frac{\opl(p)}{-\opl(0)}$ is identical to $\opll$ on $[0,1]$. In another words, 
$\opl=\kappa\opll$ with $\kappa=-\opl(0)>0$. 
\end{proof} 
\begin{remark}\label{remarkmain3}
As the lower bound $\overline{R}'\geq 0$ is satisfied by the constraint \eqref{compensation2}, the optimal retention $\overline{R}$ given by \eqref{optimalretention} is moral-hazard-free if and only if $ \overline{R}'\leq 1$ a.e., namely 
\[\frac{\kappa\opll''\big(w(F_X(x))\big)w'(F_X(x))F'_X(x)}{-u''\big((u')^{-1}\big(\kappa\opll'\big(w(F_X(x))\big)\big)}\leq 1,\quad \mbox{a.e.}\;x\in[0,\esup X],\]
or equivalently, 
\[\kappa\opll''(p)+\hbar(p)u''\big((u')^{-1}\big(\kappa\opll'(p)\big)\big)\leq 0, \quad\mbox{a.e.}\;p\in [w(\ms),1].\]
\end{remark}

\begin{remark}
As we only have one side constraint for $R'$ in this model, the related ODE becomes a single-obstacle problem. Furthermore, since the operator in the ODE is linear, we can solve it explicitly. 
\end{remark}

\subsection{PO contracts with monotone retentions}
If we only require the compensations to be monotone but put no constraint on the retentions, that is, 
\begin{align*} \setc&=\Big\{I:[0,\infty)\to [0,\infty)\; \big|\; \mbox{$I$ is absolutely }\nn\\
&\qquad\quad~\mbox{ continuous with $I(0)=0$ and $I'\geq 0$ a.e..}\Big\}.
\end{align*} 
Then the corresponding set of retentions becomes 
\begin{align} \label{compensation3}
\setr&=\Big\{R:[0,\infty)\to [0,\infty)\; \big|\; \mbox{$R$ is absolutely }\nn\\
&\qquad\quad~\mbox{ continuous with $R(0)=0$ and $ R'\leq 1$ a.e.}\Big\}.
\end{align} 
In this case the retentions in $\setr$ may not be non-decreasing.

\begin{thm}\label{main4}
If the set of retentions $\setr$ in problem \eqref{opi} is replaced by \eqref{compensation3}, then a retention $\overline{R}\in\setr$ is optimal to problem \eqref{opi} if and only if it can be written as 
\begin{align} \label{optimalretention2}
\overline{R}(x)=(u')^{-1}\big(\opl'(0)\big)-(u')^{-1}\big(\opl'\big(w(F_X(x))\big)\big)
\end{align} 
for some $\opl\in C^{2-}([0,1])$ such that 
\begin{equation}\label{vi008}
\begin{cases}
\opl''(p)=0, \hspace{246pt}\mbox{a.e.}\;p\in[0, w(\ms)], \bigskip \\
\max\Big\{\opl''(p)+\hbar(p)u''\big((u')^{-1}\big(\opl'(p)\big)\big), \;\opl(0) (1+\theta)(1-\inw(p))-\opl(p)\Big\}=0,\\\hspace{288pt}\quad\mbox{a.e.}\;p\in(w(\ms),1], \medskip\\
\opl(1)=0.
\end{cases}
\end{equation} 

\end{thm}
\begin{proof}
In this case, the requirement $R'\leq 1$ in \eqref{compensation3} leads to $Q'(p)\leq \hbar(p)$ for $p \in(w(\ms),1]$. Consequently, the optimality condition \eqref{optimalcondition1-2}
implies $\opl(p)\geq \opl(0) (1+\theta)(1-\inw(p))$ and \begin{align*}
\barq'(p)=\begin{cases}
\hbar(p), &\quad\text{if } \opl(p)>\opl(0) (1+\theta)(1-\inw(p));\\
(-\infty,\hbar(p)], &\quad\text{if } \opl(p)=\opl(0) (1+\theta)(1-\inw(p)),
\end{cases}\; \mbox{ for a.e. $p\in(w(\ms),1]$,}
\end{align*} 
which is clearly equivalent to 
\begin{align*}
\max\Big\{\barq'(p)-\hbar(p),\opl(0) (1+\theta)(1-\inw(p))-\opl(p)\Big\}=0, \quad\mbox{a.e.}\;p\in(w(\ms),1].
\end{align*} 
This leads to the ODE \eqref{vi008} for $\opl$.
\end{proof} 

\begin{remark}\label{remarkmain4}
As the upper bound $\overline{R}'\leq 1$ is satisfied by the constraint \eqref{compensation3}, the optimal retention $\overline{R}$ given by \eqref{optimalretention2} is moral-hazard-free if and only if $\overline{R}'\geq 0$. The latter is equivalent to the solution $\opl$ to \eqref{vi008} is convex. 
\end{remark}

\begin{coro}
If the probability weighting function $w$ is concave (such as no distortion case $w(p)\equiv p$), then the optimal retention $\overline{R}$ given by \eqref{optimalretention2} is moral-hazard-free. 
\end{coro}
\begin{proof}
By \citeremark{remarkmain4} it suffices to show the solution $\opl$ to \eqref{vi008} is convex. Suppose on the contrary, the region $\{p\in (w(\ms),1): \opl''(p)<0 \}$ was not empty. In this region, we would have $\opl''(p)+\hbar(p)u''\big((u')^{-1}\big(\opl'(p)\big)\big)<0$; and consequently, $\opl(p)=\opl(0) (1+\theta)(1-\inw(p))$ by \eqref{vi008}. This would imply $-\opl(0) (1+\theta)\inw''(p)=\opl''(p)<0$. As $\opl(0)<0$, we conclude $\inw''(p)<0$ which contradicts the concavity of $w$. 
\end{proof}

\begin{remark}
The ODE \eqref{vi008} is still a single-obstacle problem as the previous model, but its operator is nonlinear in general so we cannot provide an explicit solution. When $u$ is an exponential utility, the operator, however, reduces to a linear one (see \eqref{vi002-1}). One can still show that $\opl$ after change of variable is the convex envelope of some known function. We encourage the interested reader to write down the details. 
\end{remark}

\section{Concluding remarks}\label{conclusion}
In this paper, we present a novel approach to computing all the PO moral-hazard-free insurance contracts under the RDU theory. Similar to \cite{XZ16,X16}, the approach also works for problems under some other behavioral finance theories. For instance, one could use our method to consider the loss and gain parts in the cumulative prospect theory model, separately, and then combine them to get the optimal solution. Our method also allows us to find all the PO moral-hazard-free insurance contracts with general lower and/or upper bounds on the derivatives of the retentions and/or the compensations. 
\par
In our model, we considered the expected value premium principle for the insurer. It is possible to generalize our method to cope with models with other premium principles such as the variance premium principle and Wang's premium principle for the insurer. In these cases, the problem becomes more challenging. We leave them for our future research. 
\bigskip

\textbf{Acknowledgments.} The author is grateful to the Co-Editor Professor Jaksa Cvitanic and two anonymous referees for their constructive comments and suggestions which have helped to significantly improve the paper of the two previous versions, and to Mr. Jing Peng for help in calculating the numerical example in \citesec{numericalexample}.
The author is also grateful for comments from conference participants at The 12th AIMS Conference on Dynamical Systems, Differential Equations and Applications in Taipei, 2018 Conference in Memory of Professor Xunjing Li in Changchun, SCFM 2018 in Qingdao, 2018 International Conference on Mathematical Finance \& Symposium on the Role of Mathematical Finance on FinTech Business in Daejeon, 2018 Stochastic Analysis, Stochastic Control and New Developments in Weihai, The 8th Annual Conference of Financial Engineering \& Financial Risk Management Branch of OR Society of China in Xi'an, 2018 Advanced Methods in Mathematical Finance in Angers, The Sixth Asian Quantitative Finance Conference in Guangzhou, The First Conference on Actuarial Science and Applications in Shanghai.



\end{document}